%% file: main.tex
\RequirePackage{snapshot}
\documentclass[acmsmall,screen]{acmart}

\usepackage{booktabs}   
\usepackage{subcaption} 

\usepackage{ifthen}
\newcommand{\isTechReport}{true} 
\newcommand\includeTechReport[1]{%
  \ifthenelse{\equal{\isTechReport}{true}}
    {{#1}}
    {\ignorespaces}
\xspace}

\usepackage{tikz}
\usetikzlibrary{positioning}
\usepackage{pgfplots}
\pgfplotsset{compat=1.12}
\usepackage{pgfplotstable}

\usepackage{amsmath}
\usepackage{amssymb}
\usepackage{xspace}
\usepackage{booktabs}
\usepackage{listings}
\lstnewenvironment{code}{
\lstset{
  language=Caml,
  basicstyle=\ttfamily,
  keywordstyle=\ttfamily,
}}
{}

\lstnewenvironment{ecode}{
\lstset{
  language=Caml,
  basicstyle=\ttfamily,
  keywordstyle=\ttfamily,
  numbers=left,
  frame=leftline,
  xleftmargin=7.5mm,
  moredelim=[is][\bfseries]{==}{==},
  moredelim=[is][\underbar]{__}{__},
  moredelim=[is][\bfseries\underbar]{_=}{=_},
  escapeinside={(*@}{@*)},
}}
{}

\lstMakeShortInline{|}

\usepackage{tcolorbox}

\definecolor{tree}{HTML}{8dd3c7}
\definecolor{sherrloc}{HTML}{FFFFB3}
\definecolor{fix}{HTML}{bebada}

\tcbset{every box/.style={on line,colframe=black,size=fbox,boxrule=1pt}}
\newtcbox{\hlTree}{colback=tree,toprule=0pt,bottomrule=0pt}
\newtcbox{\hlSherrloc}{colback=sherrloc,leftrule=0pt,rightrule=0pt}
\newtcbox{\hlFix}{colback=fix!50}

\usepackage{commands}

\setcopyright{rightsretained}
\acmPrice{}
\acmDOI{10.1145/3138818}
\acmYear{2017}
\acmJournal{PACMPL}
\acmVolume{1}
\acmNumber{OOPSLA}
\acmArticle{60}
\acmMonth{10}

\acmBadgeL[https://github.com/ucsd-progsys/nate]{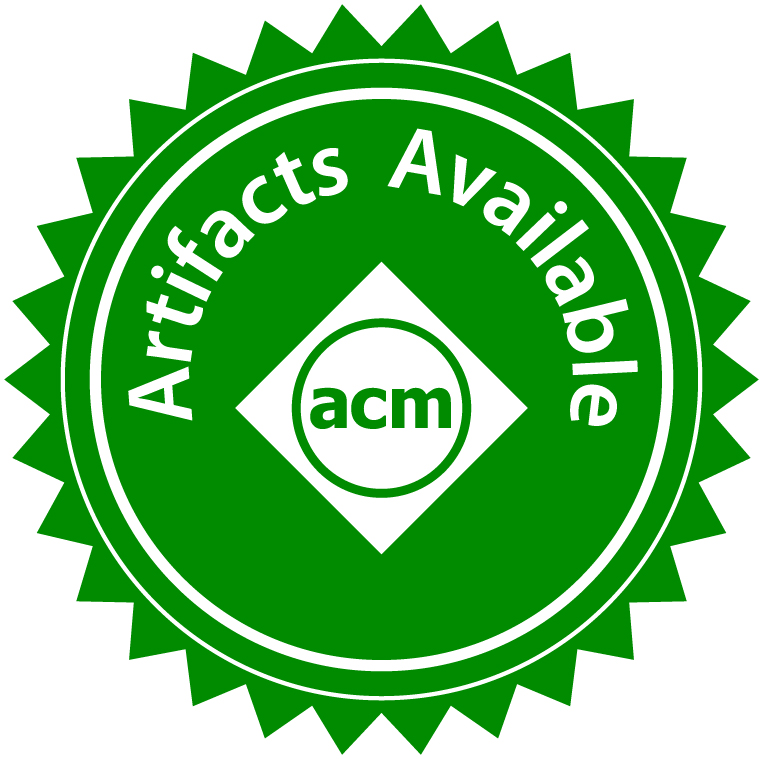}
\acmBadgeR[https://www.acm.org/publications/policies/artifact-review-badging]{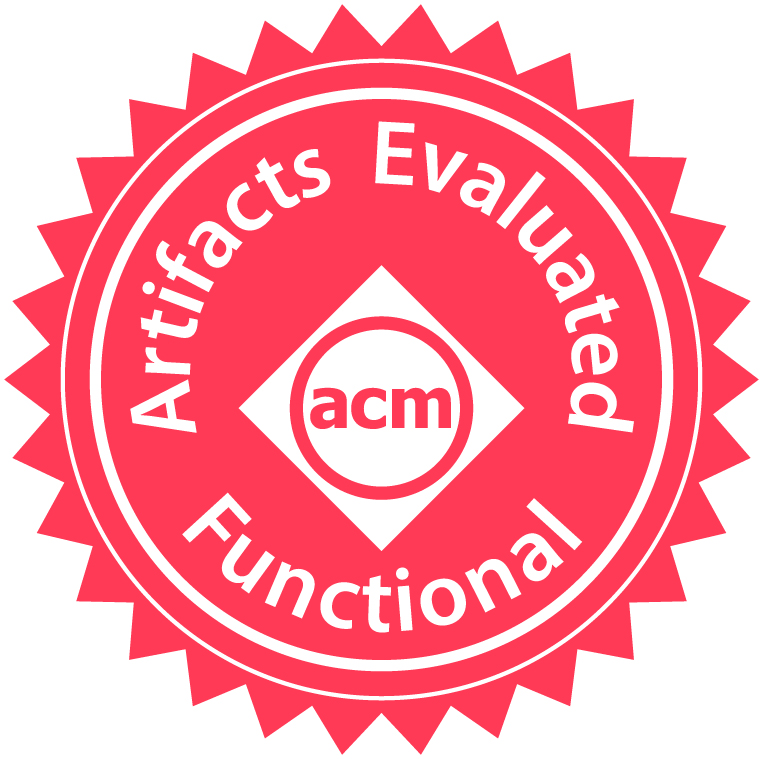}

\begin{document}

\title{Learning to Blame}

\titlenote{This work was supported by NSF grants CCF-1422471, CCF-1223850,
  CCF-1218344, CCF-1116289, CCF-0954024, IIS-1617157, Air Force grant
  FA8750-15-2-0075, and a gift from Microsoft Research.}

\subtitle{Localizing Novice Type Errors with Data-Driven Diagnosis}



\author{Eric L. Seidel}
\orcid{0000-0002-2529-7790}               
\affiliation{
  \department{Department of Computer Science}              
  \institution{University of California, San Diego}            
  \city{La Jolla}
  \state{CA}
  \country{USA}
}
\email{eseidel@cs.ucsd.edu}          

\author{Huma Sibghat}
\affiliation{
  \department{Department of Computer Science}              
  \institution{University of California, San Diego}            
  \city{La Jolla}
  \state{CA}
  \country{USA}
}
\email{hsibghat@cs.ucsd.edu}          

\author{Kamalika Chaudhuri}
\affiliation{
  \department{Department of Computer Science}              
  \institution{University of California, San Diego}            
  \city{La Jolla}
  \state{CA}
  \country{USA}
}
\email{kamalika@cs.ucsd.edu}          

\author{Westley Weimer}
\affiliation{
  \department{Department of Computer Science}              
  \institution{University of Virginia}            
  \country{USA}
}
\email{weimer@cs.virginia.edu}          

\author{Ranjit Jhala}
\affiliation{
  \department{Department of Computer Science}              
  \institution{University of California, San Diego}            
  \city{La Jolla}
  \state{CA}
  \country{USA}
}
\email{jhala@cs.ucsd.edu}          



\input{abstract}



\maketitle

\input{intro2}
\input{overview}
\input{learning}
\input{evaluation}
\input{discussion}
\input{related}
\input{conclusion}

\section*{Acknowledgments}
We thank Jiani Huang and Yijun Zhang for helping analyze the student
interaction traces, Alexander Bakst and Valentin Robert for assisting
with the user study, and the anonymous reviewers for their insightful
feedback.

\bibliography{main}

\includeTechReport{
\clearpage
\appendix
\section{User Study}
\label{sec:user-study-exams}
\subsection{Version A}
\noindent\fbox{\includegraphics[width=0.98\linewidth,page=1]{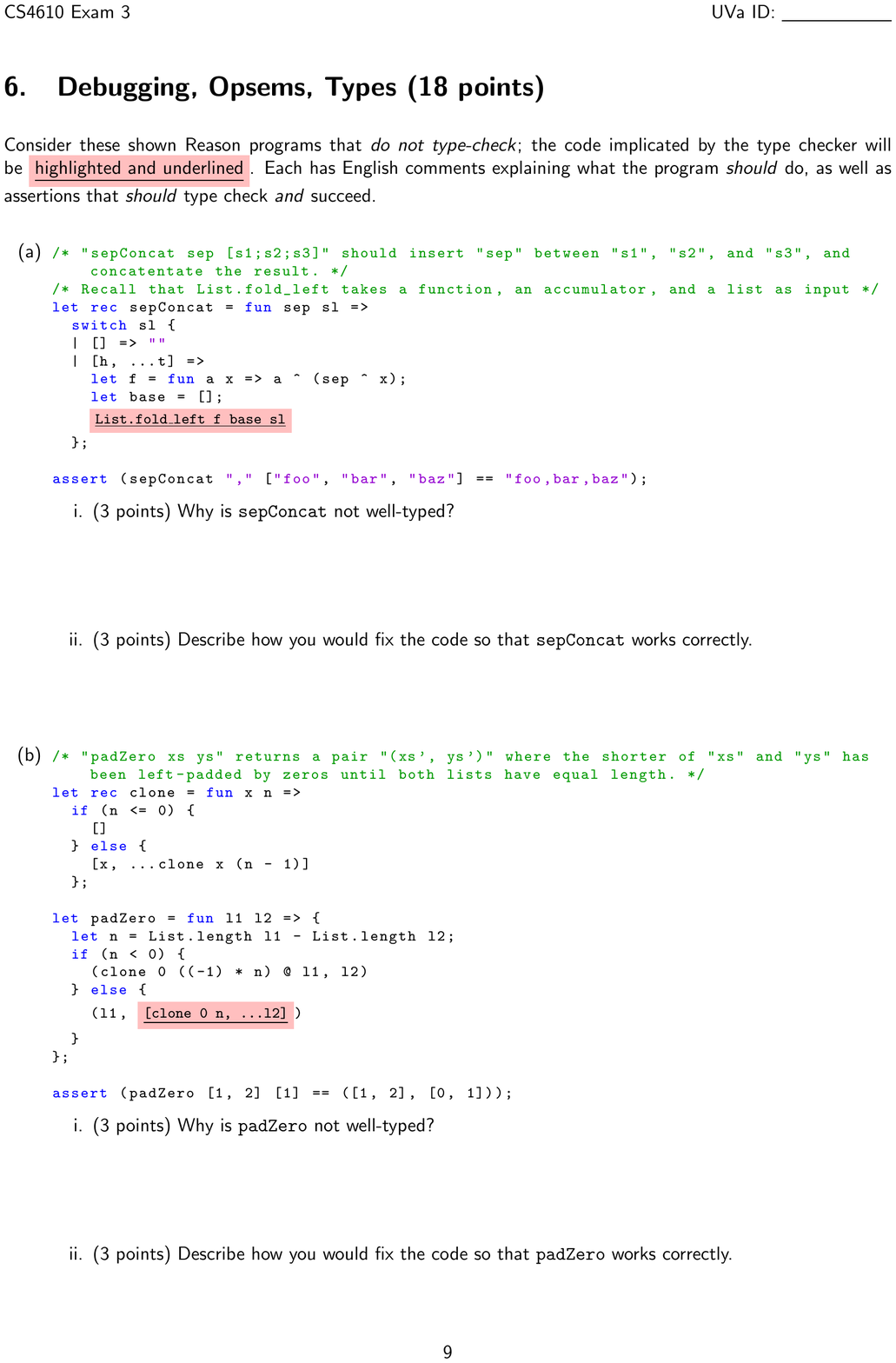}}
\newpage
\noindent\fbox{\includegraphics[width=0.98\linewidth,page=2]{study/study_a.pdf}}
\newpage
\subsection{Version B}
\noindent\fbox{\includegraphics[width=0.98\linewidth,page=1]{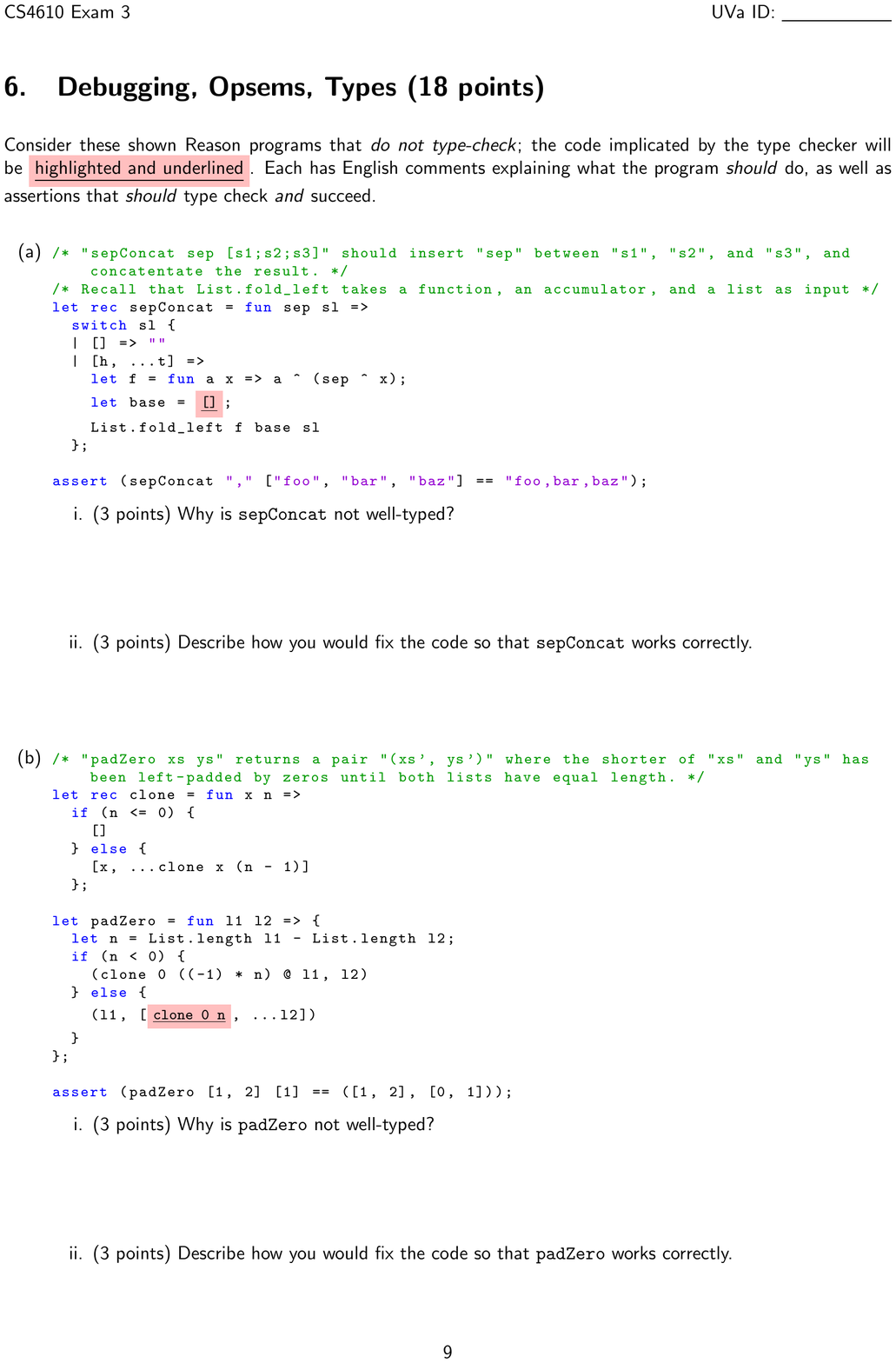}}
\newpage
\noindent\fbox{\includegraphics[width=0.98\linewidth,page=2]{study/study_b.pdf}}
}

\end{document}

%% file: abstract.tex
\begin{abstract}
Localizing type errors is challenging in
languages with global type inference, as
the type checker must make assumptions
about what the programmer intended to do.
We introduce \toolname, a \emph{data-driven}
approach to error localization based on
supervised learning.
\toolname analyzes a large corpus
of training data --- pairs of ill-typed
programs and their ``fixed'' versions ---
to automatically \emph{learn a model}
of where the error is most likely
to be found.
Given a new ill-typed program,
\toolname executes the model to
generate a list of potential blame
assignments ranked by likelihood.
We evaluate \toolname by comparing its
precision to the state of the art
on a set of over 5,000 ill-typed \ocaml
programs drawn from two instances of an
introductory programming course.
We show that when the top-ranked blame assignment
is considered, \toolname's data-driven
model is able to correctly predict
the exact sub-expression that should
be changed \HiddenFhTopOne\% of the time, 
\ToolnameWinOcaml points higher than \ocaml and
\ToolnameWinSherrloc points higher than the state-of-the-art
\sherrloc tool.
Furthermore, \toolname's accuracy surpasses
\HiddenFhTopTwo\% when we consider the top \emph{two}
locations and reaches \HiddenFhTopThree\% if we consider
the top \emph{three}.
\end{abstract}

%% file: intro2.tex
\section{\textbf{Introduction}}
\label{sec:introduction}

%
Types are awesome.
%
Languages like \ocaml and \haskell make
the value-proposition for types even more
appealing by using constraints to automatically
synthesize the types for all program terms
without troubling the programmer for any
annotations.
Unfortunately, this automation has come at a price.
Type annotations signify
the programmer's intent and help to correctly
blame the erroneous sub-term when the code is
ill-typed.
In the absence of such signifiers, automatic
type inference algorithms are prone to report
type errors far from their source
\citep{Wand1986-nw}.
While this can seem like a minor annoyance to
veteran programmers, \citet{Joosten1993-yx} have found
that novices often focus their attention on the \emph{location}
reported and disregard the \emph{message}.

\mypara{Localizing Errors}
Several recent papers have proposed ways
to improve feedback via error \emph{localization}.
At a high-level these techniques analyze
the set of typing constraints to find
the minimum (weighted) subset that,
if removed, would make the constraints
satisfiable and hence, assertion-safe~\citep{Jose:2011}
or well-typed~\citep{Zhang2014-lv,Loncaric2016-uk,Chen2014-gd,Pavlinovic2014-mr}.
The finger of blame is then pointed at the
sub-terms that yielded those constraints.
This minimum-weight approach suffers
from two drawbacks.
First, they are not \emph{extensible}:
the constraint languages and algorithms for computing
the minimum weighted subset must be
designed afresh for different kinds
of type systems and constraints \citep{Loncaric2016-uk}.
Second, and perhaps most importantly,
they are not \emph{adaptable}: the
weights are fixed in an ad-hoc fashion, based on the
\emph{analysis designer's} notion
of what kinds of errors are more
likely, rather than
adapting to the kinds of mistakes
programmers actually make in practice.

\mypara{A Data-Driven Approach}
In this paper, we introduce \toolname
\footnote{``Numeric Analysis of Type Errors''; any resemblance to persons living or dead is purely coincidental.}
a \emph{data-driven} approach to error
localization based on supervised
learning (see \citealt{Kotsiantis2007-pj} for a survey).
\toolname analyzes a large corpus
of training data --- pairs of ill-typed
programs and their subsequent fixes ---
to automatically \emph{learn a model}
of where errors are most likely to
be found.
Given a new ill-typed program,
\toolname simply executes the model
to generate a list of potential
blame assignments ranked by likelihood.
We evaluate \toolname by comparing its
precision against the state-of-the-art
on a set of over 5,000 ill-typed \ocaml
programs drawn from two instances of an
introductory programming course.
We show that, when restricted to a
\emph{single} prediction, \toolname's data-driven
model is able to correctly predict
the exact sub-expression that should
be changed \HiddenFhTopOne\% of the time,
\ToolnameWinOcaml points higher than \ocaml and
\ToolnameWinSherrloc points higher than the state-of-the-art
\sherrloc tool.
Furthermore, \toolname's accuracy surpasses
\HiddenFhTopTwo\% when we consider the top \emph{two}
locations and reaches \HiddenFhTopThree\% if we consider
the top \emph{three}.
We achieve these advances by identifying
and then solving three key challenges.

\mypara{Challenge 1: Acquiring Labeled Programs}
The first challenge for supervised learning
is to acquire a corpus of training data, in our setting
a set of ill-typed programs \emph{labeled}
with the exact sub-terms that are the actual
cause of the type error.
Prior work has often enlisted expert users
to manually judge ill-typed programs and
determine the ``correct'' fix
\citep[\eg][]{Lerner2007-dt,Loncaric2016-uk},
but this approach does not scale well to
a dataset large enough to support machine
learning.
Worse, while expert users have intimate
knowledge of the type system, they may
have a blind spot with regards to the
kinds of mistakes novices make, and
cannot know in general what novice users
intended.

Our \emph{first contribution} (\autoref{sec:overview})
is a set of more than 5,000 labeled programs \citep{Seidel2017-ko},
giving us an accurate ground truth of
the kinds of errors and the (locations
of the) fixes that novices make in
practice.
We obtain this set by observing that
software development is an iterative
process; programmers eventually
fix their own ill-typed programs,
perhaps after multiple incorrect
exploratory attempts.
To exploit this observation we instrumented
the \ocaml compiler to collect fine-grained
traces of student interactions over two instances
of an undergraduate Programming Languages
course
\begin{anonsuppress}
at UC San Diego (IRB \#140608).
\end{anonsuppress}
\begin{noanonsuppress}
at AUTHOR's INSTITUTION (IRB HIDDEN).
\end{noanonsuppress}
We then post-process the resulting time-series
of programs submitted to the \ocaml compiler into
a set of pairs of ill-typed programs and their
subsequent \emph{fixes}, the first (type-)~correct
program in the trace suffix.
Finally, we compute the blame labels using a
\emph{tree-diff} between the two terms to find
the exact sub-terms that changed in the fix.

\mypara{Challenge 2: Modeling Programs as Vectors}
Modern supervised learning algorithms work on 
\emph{feature vectors}: real-valued points in an
$n$-dimensional space. While there are standard
techniques for computing such
vectors for documents, images, and sound (respectively
word-counts, pixel-values, and frequencies),
there are no similarly standard representations for
programs.

Our \emph{second contribution} (\autoref{sec:learning})
solves this problem with a simple, yet expressive, representation called
a \emph{Bag-of-Abstracted-Terms} (BOAT) wherein
each program is represented by the \emph{bag}
or multiset of (sub-)~terms that appears inside
it; and further, each (sub-)~term is \emph{abstracted}
as a feature vector comprising the numeric values
returned by \emph{feature abstraction} functions
applied to the term.
%
%
We can even recover \emph{contextual} information
from the parent and child terms by
\emph{concatenating} the feature vectors of each term
with those of its parent and children
(within a fixed window).
We have found this representation to be particularly
convenient as it gives us flexibility in modeling the
syntactic and semantic structure of programs while
retaining compatibility with off-the-shelf classifiers,
in contrast to, \eg, \citet{Raychev2015-jg}, who had
to develop their own variants of classifiers to obtain
their results.

\mypara{Challenge 3: Training Precise Classifiers}
Finally, the last and most important challenge is to
use our BOAT representation to train classifiers that
are capable of \emph{precisely} pinpointing the errors
in real programs.
The key here is to find the right set of feature
abstractions to model type errors,
and classification algorithms that
lead to precise blame assignments.
Fortunately, our BOAT model allows us a great deal of
latitude in our choice of features.
We can use abstraction functions to capture different
aspects of a term ranging from
syntactic features (\eg is-a-data-constructor, is-a-literal,
is-an-arithmetic-operation, is-a-function-application, \etc),
to semantic features captured by the type system (\eg is-a-list,
is-an-integer, is-a-function, \etc).
We can similarly model the blame labels with a simple feature
abstraction (\eg is-changed-in-fix).

Our \emph{third contribution} (\autoref{sec:evaluation})
is a systematic evaluation of our data-driven approach
using different classes of features like the above, and
with four different classification algorithms: logistic
regression, decision trees, random forests, and neural networks.
We find that \toolname's models \emph{generalize} well
between instances of the same undergraduate course, outperforming
the state of the art by at least \ToolnameWinSherrloc
percentage points at predicting the source of a type error.
We also investigate which features and classifiers
are most effective at localizing type errors, and
empirically characterize the importance of different
feature sets.
%
In particular, we find that while machine learning
over syntactic features of each term in isolation
performs worse than existing
purely constraint-based approaches (\eg \ocaml, \sherrloc),
augmenting the data with a single feature corresponding to
the \emph{type error slice} \citep{Tip2001-qp} brings our
classifiers up to par with the state-of-the-art,
and further augmenting the data with \emph{contextual}
features allows our classifiers to outperform
the state-of-the-art by \ToolnameWinSherrloc percentage points.

Thus, by combining modern statistical methods
with domain-specific feature engineering, \toolname
opens the door to a new data-driven path to
precise error localization.
In the future, we could \emph{extend}
\toolname to new languages or forms
of correctness checks by swapping in
a different set of feature abstraction
functions.
Furthermore, our data-driven approach
allows \toolname to \emph{adapt} to
the kinds of errors that programmers
(in particular novices, who are in greatest
need of precise feedback) actually make
rather than hardwiring the biases of
compiler authors who, by dint of their
training and experience, may suffer from
blind spots with regards to such problems.
In contrast, our results show that \toolname's
data-driven diagnosis can be an effective
technique for localizing errors by collectively
learning from past mistakes.


%% file: overview.tex
\mysection{Overview}\label{sec:overview}


\begin{figure}[t!]
\small
\begin{minipage}{0.45\linewidth}
\begin{ecode}
  let rec sumList xs =
    match xs with
    | []   -> (*@\hlTree{\hlSherrloc{[]}}@*)
    | h::t -> h (*@\hlTree{+}@*) (*@\hlFix{sumList t}@*)
\end{ecode}
\end{minipage}
\begin{minipage}{0.49\linewidth}
\begin{verbatim}
File "sumList.ml", line 4, characters 16-25:
  This expression has type 'a list
  but an expression was expected of type int
\end{verbatim}
\end{minipage}
\caption{(left) An ill-typed \ocaml program that should sum the elements of a
  list, with highlights indicating three possible blame assignments based on:
  (1) the \hlFix{\ocaml} compiler;
  (2) the \hlSherrloc{fix} made by the programmer; and
  (3) \hlTree{minimizing} the number of edits required.
  %
  %
  (right) The error reported by \ocaml.}
\label{fig:sumList}
\end{figure}


Let us start with an overview of \toolname's
approach to localizing type errors by
collectively learning from the mistakes
programmers actually make.

\mypara{The Problem}
Consider the |sumList| program in
\autoref{fig:sumList}, written by
a student in an undergraduate
Programming Languages course.
The program is meant to add up the
integers in a list, but the student
has accidentally given the empty
list as the base case, rather than |0|.
The \ocaml compiler collects typing
constraints as it traverses the program,
and reports an error the moment it finds
an inconsistent constraint.
In this case it blames the recursive call
to |sumList|, complaining that |sumList|
returns a |list| while an |int| was
expected by the |+| operator.
This \emph{blame} assignment is inconsistent
with the programmer's intention and may
not help the novice understand the error.

It may appear obvious to the reader that
|[]| is the correct expression to blame,
but how is a type checker to know that?
Indeed, 
recent techniques like
\sherrloc and \mycroft
\citep{Zhang2014-lv,Loncaric2016-uk,Pavlinovic2014-mr}
fail to distinguish between
the |[]| and |+| expressions
in \autoref{fig:sumList};
it would be equally valid
to blame \emph{either}
of them alone.
The |[]| on line 3 could be changed to |0|,
or the |+| on line 4 could be changed to
either |@| (list append) or |::|, all of
which would give type-correct programs.
Thus, these state-of-the-art techniques
are forced to either blame \emph{both}
locations, or choose one \emph{arbitrarily}.

\mypara{Solution: Localization via Supervised Classification}
Our approach is to view error localization as a
\emph{supervised classification}
problem~\citep{Kotsiantis2007-pj}.
A \emph{classification} problem entails learning
a function that maps \emph{inputs} to a discrete
set of output \emph{labels} (in contrast to 
\emph{regression}, where the output is typically
a real number).
A \emph{supervised} learning problem is one where
we are given a \emph{training set} where the
inputs and labels are known, and the task is to
learn a function that accurately maps the inputs
to output labels and \emph{generalizes} to new,
yet-unseen inputs.
To realize the above approach for error localization
as a practical tool, we have to solve four sub-problems.
\begin{enumerate}
  \item How can we acquire a \emph{training set} of
        blame-labeled ill-typed programs?

  \item How can we \emph{represent} blame-labeled programs
        in a format amenable to machine learning?

  \item How can we find \emph{features} that yield predictive
        models?

  \item How can we use the models to give localized
        \emph{feedback} to the programmer?
\end{enumerate}

\mysubsection{Step 1: Acquiring a Blame-Labeled Training Set}

The first step is to gather a training
set of ill-typed programs, where each
erroneous sub-term is explicitly labeled.
Prior work has often enlisted
expert users to curate a set of
ill-typed programs and then
\emph{manually} determine the
correct fix~\citep[\eg][]{Lerner2007-dt,Loncaric2016-uk}.
This method is suitable for
\emph{evaluating} the quality
of a localization (or repair)
algorithm on a small number
(\eg 10s--100s) of programs.
However, in general it requires
a great deal of effort for the
expert to divine the original
programmer's intentions.
Consequently, is difficult to
scale the expert-labeling to
yield a dataset large enough
(\eg 1000s of programs)
to facilitate machine learning.
More importantly, this approach
fails to capture the \emph{frequency}
with which errors occur in practice.

\mypara{Solution: Interaction Traces}
We solve both the scale and
frequency problems by instead
extracting blame-labeled data sets
from \emph{interaction traces}.
Software development is an iterative process.
Programmers, perhaps after a lengthy (and
sometimes frustrating) back-and-forth with
the type checker, eventually end up fixing
their own programs.
Thus, we instrumented
the \ocaml compiler to record
this conversation, \ie record the sequence
of programs submitted by each programmer and
whether or not it was deemed type-correct.
For each ill-typed program in
a particular programmer's trace,
we find the \emph{first subsequent}
program in the trace that type checks
and declare it to be the fixed version.
From this pair of an ill-typed program
and its fix, we can extract a \emph{diff}
of the abstract syntax trees, and then assign
the blame labels to the \emph{smallest}
sub-tree in the diff.


\mypara{Example}
Suppose our student
fixed the |sumList| program in
\autoref{fig:sumList} by replacing
|[]| with |0|, the diff would
include only the |[]| expression.
Thus we would determine that the
|[]| expression (and \emph{not} the
|+| or the recursive call |sumList t|)
is to blame.


\mysubsection{Step 2: Representing Programs as Vectors}

Next, we must find a way to translate
highly structured and variable sized
\emph{programs} into fixed size
$n$-dimensional numeric \emph{vectors}
that are needed for supervised
classification.
While the Programming Languages literature is full
of different program
representations, from raw
token streams to
richly-structured
abstract syntax trees (AST) or
control-flow graphs, it is
unclear how to embed the
above into a vector space.
Furthermore, it is unclear whether
recent program representations that
are amenable to one learning task,
\eg code completion \citep{Devanbu:2012,Raychev:2014}
or decompilation \citep{Raychev2015-jg,Bielik2016-br}
are suitable for our problem of
assigning blame for type errors.

\mypara{Solution: Bags-of-Abstracted-Terms}
We present a new representation of programs
that draws inspiration from the theory of
abstract interpretation \citep{CousotCousot77}.
Our representation is parameterized by a
set of \emph{feature abstraction} functions,
(abbreviated to feature abstractions)
$f_1, \ldots, f_n$, that map terms to a
numeric value (or just $\{0, 1\}$ to
encode a boolean property).
Given a set of feature abstractions, we
can represent a single program's AST as
a \emph{bag-of-abstracted-terms} (BOAT)
by:
(1)~decomposing the AST (term) $t$ into
    a \emph{bag} of its constituent sub-trees
    (terms) $\{t_1,\ldots,t_m\}$; and then
(2)~representing each sub-term $t_i$
    with the $n$-dimensional
    vector $[f_1(t_i),\ldots, f_n(t_i)]$.
Working with ASTs is a natural choice
as type-checkers operate on the same representation.

\mypara{Modeling Contexts}
Each expression occurs in some surrounding
\emph{context}, and we would like the
classifier to be able make decisions based
on the context as well.
The context is particularly important
for our task as each expression
imposes typing constraints on its
neighbors.
For example, a |+| operator tells
the type checker that both \emph{children}
must have type |int| and that the \emph{parent}
must accept an |int|.
Similarly, if the student wrote
|h sumList t| \ie forgot the |+|,
we might wish to blame the application
rather than |h| because |h|
\emph{does not} have a function type.
The BOAT representation makes it
easy to incorporate contexts: we
simply \emph{concatenate} each
term's feature vector with the
\emph{contextual features} of
its parent and children.



\mysubsection{Step 3: Feature Discovery}

Next, we must find a \emph{good}
set of features, that is, a set
of features that yields predictive
models. Our BOAT representation
enables an iterative solution
by starting with
a simple set of features, and
then repeatedly adding more
and more to capture important
aspects needed to improve precision.
Our set of feature abstractions
captures the
\emph{syntax}, \emph{types}, and
\emph{context} of each expression.

\mypara{Syntax and Type Features}
We start by observing that
at the very least, the
classifier should be able
to distinguish between the
|[]| and |+| expressions
in \autoref{fig:sumList}
because they represent
different \emph{syntactic}
expression forms.
We model this by
introducing feature
abstractions of the form
is-|[]|, is-|+|, \etc, for
each of a fixed number of
expression forms.
Modeling the syntactic class of an
expression gives the classifier a
basic notion of the relative
frequency of blame assignment
for the various program elements,
\ie perhaps |[]| is
\emph{empirically} more
likely to be blamed than |+|.
Similarly, we can model
the \emph{type} of each
sub-expression with features
of the form is-|int|, is-|bool|, \etc.
We will discuss handling
arbitrary, user-defined types
in~\autoref{sec:discussion}.

\mypara{Contextual Features: Error Slices}
Our contextual features include the
syntactic class of the neighboring
expressions and their inferred types
(when available).
However, we have found that
the most important contextual
signal is whether or not the
expression occurs in
a minimal type error slice
\citep{Tip2001-qp,Haack2003-vc}
which includes \emph{a minimal subset}
of all expressions that are
necessary for the error to manifest.
(That is, replacing any subterm
with |undefined| or |assert false|
would yield a well-typed program.)
We propose to use type error slices
to communicate to the classifier
which expressions could
\emph{potentially} be blamed --- a
change to an expression outside of
the minimal slice cannot possibly
fix the type error.
We empirically demonstrate that
the type error slice is so
important (\autoref{sec:feature-utility})
that it is actually beneficial to
automatically discard expressions
that are not part of the slice,
rather than letting the classifier
learn to do so.
Indeed, this domain-specific
insight is crucial for learning
classifiers that significantly
outperform the state-of-the-art.

\mypara{Example}
When \toolname is tasked with localizing
the error in the example program of \autoref{fig:sumList},
the |[]| and |+| sub-terms will each be given
their own feature vector, and we will ask the
classifier to predict for each \emph{independently}
whether it should be blamed.
\autoref{tab:sumList} lists some
of the sub-expressions of the example
from \autoref{fig:sumList}, and their
corresponding feature vectors.

\input{feature-table}


\mysubsection{Step 4: Generating Feedback}

Finally, having trained a classifier
using the labeled data set, we need to use
it to help users localize type errors in
their programs.
The classifier tells us whether or not
a sub-term \emph{should be}
blamed (\ie has the blame label) but this
is not yet particularly suitable as
\emph{user feedback}.
A recent survey of developers by
\citet{Kochhar2016-oc} found that
developers are unlikely to examine
more than around five potentially
erroneous locations before falling
back to manual debugging.
Thus, we should limit our predictions
to a select few to be presented to
the user.

\mypara{Solution: Rank Locations by Confidence}
Fortunately, many machine learning
classifiers produce not only a predicted
label, but also a metric that can be
interpreted as the classifier's
\emph{confidence} in its prediction.
Thus, we \emph{rank} each expression
by the classifier's confidence that
it should be blamed, and present only
the top-$k$ predictions to the
user (in practice $k=3$).
The use of ranking to report the
results of an analysis is
popular in other problem domains
\citep[see, \eg][]{Kremenek2003-ck};
we focus explicitly on the use of
data-driven machine learning
confidence as a ranking source.
In \autoref{sec:evaluation} we show
that \toolname's ranking approach
yields a high-precision localizer:
when the top three locations are considered,
at least one matches an actual student fix
\HiddenFhTopThree\% of the time.



%% file: feature-table.tex
\begin{table}[ht]
\caption{Example Feature Vectors}\label{tab:sumList}
\begin{tabular}{lrrrrrr}
\toprule
\textbf{Expression}
  & \IsNil & \IsCaseListP & \ExprSize
  & \HasTypeIntCOne & \HasTypeList & \InSlice \\
\midrule
|[]|
  & 1 & 1 & 1 & 0 & 1 & 1 \\
|hd + sumList tl|
  & 0 & 1 & 5 & 1 & 0 & 1 \\
|sumList tl|
  & 0 & 0 & 3 & 0 & 1 & 1 \\
|tl|
  & 0 & 0 & 1 & 0 & 1 & 0 \\
\bottomrule
\end{tabular}
\bigskip
\caption*{A selection of the features we would extract from the
\lstinline!sumList! program in \autoref{fig:sumList}. A feature is
considered \emph{enabled} if it has a non-zero value, and
\emph{disabled} otherwise. A ``-P'' suffix indicates that the feature
describes the parent of the current expression, a ``-C$n$'' suffix
indicates that the feature describes the $n$-th (left-to-right) child of
the current expression.  Note that, since we rely on a partial typing
derivation, we are subject to the well-known traversal bias and label
the expression \lstinline!sumList tl! as having type
$\tlist{\cdot}$. The model will have to learn to correct for this bias.}
\end{table}

%% file: learning.tex
\mysection{Learning to Blame}
\label{sec:learning}
In this section, we describe our approach to localizing type errors, in the
context
of \lang (\autoref{fig:syntax}), a simple lambda calculus with integers,
booleans, pairs, and lists.
Our goal is to instantiate the $\blamesym$ function of
\autoref{fig:api}, which takes as input a $\Model$ of type errors and an
ill-typed program $e$, and returns an ordered list of subexpressions
from $e$ paired with the confidence score $\Runit$ that they should be
blamed.

A $\Model$ is produced by $\trainsym$, which performs
supervised learning on a training set of feature vectors $\V$ and
(boolean) labels $\B$.
Once trained, we can $\evalsym$uate a $\Model$ on a new input,
producing the confidence $\Runit$ that the blame label should be
applied.
We describe multiple $\Model$s and their instantiations of
$\trainsym$ and $\evalsym$
(\autoref{sec:models}).

Of course, the $\Model$ expects feature vectors $\V$ and blame labels
$\B$, but we are given program pairs.
So our first step must be to define a suitable translation from program
pairs to feature vectors and labels, \ie we must define the
$\extractsym$ function in \autoref{fig:api}.
We model features as real-valued functions of
terms, and extract a feature vector for each \emph{subterm}
of the ill-typed program (\autoref{sec:features}).
Then we define the blame labels for the training set to be the
subexpressions that changed between the ill-typed program and its
subsequent fix, and model $\blamesym$ as a function from a program pair
to the set of expressions that changed (\autoref{sec:labels}).
The $\extractsym$ function, then, extracts $\featuresym$ from each
subexpression and computes the blamed expressions according to
$\labelsym$.


\input{syntax}

\mysubsection{Features}
\label{sec:features}


The first issue we must tackle is formulating our learning task in
machine learning terms.
We are given programs over \lang, but learning algorithms expect to work
with \emph{feature vectors} $\V$ --- vectors of real numbers, where each
column describes a particular aspect of the input.
Thus, our first task is to convert programs to feature vectors.

We choose to model a program as a \emph{set} of feature vectors, where
each element corresponds an expression in the program.
Thus, given the |sumList| program in \autoref{fig:sumList} we
would first split it into its constituent sub-expressions and then
transform each sub-expression into a single feature vector.
We group the features into five categories, using \autoref{tab:sumList}
as a running example of the feature extraction process.

\mypara{Local syntactic features}
These features describe the syntactic category of each expression $e$.
In other words, for each production of $e$ in \autoref{fig:syntax} we
introduce a feature that is enabled (set to $1$) if the expression was
built with that production, and disabled (set to $0$) otherwise.
For example, the \IsNil feature in \autoref{tab:sumList} describes
whether an expression is the empty list $\enil$.

We distinguish between matching on a list vs.\ on a pair, as this
affects the typing derivation.
We also assume that all pattern matches are well-formed --- \ie all
patterns must match on the same type.
Ill-formed match expressions would lead to a type error; however, they
are already effectively localized to the match expression itself.
We note that this is not a \emph{fundamental} limitation, and one could
easily add features that specify whether a match \emph{contains} a
particular pattern, and thus have a match expression that enables multiple
features.

\mypara{Contextual syntactic features}
These are similar to local syntactic features, but lifted to describe the
parent and children of an expression.
For example, the \IsCaseListP feature in \autoref{tab:sumList} describes
whether an expression's \emph{parent} matches on a list.
If a particular $e$ does not have children (\eg a variable $x$) or a
parent (\ie the root expression), we leave the corresponding features
disabled.
This gives us a notion of the \emph{context} in which an expression
occurs, similar to the \emph{n-grams} used in linguistic
models \citep{Hindle2012-hf,Gabel2010-el}.


\mypara{Expression size}
We also propose a feature representing the \emph{size} of each expression,
\ie how many sub-expressions does it contain?
For example, the \ExprSize feature in \autoref{tab:sumList} is set to three
for the expression |sumList tl| as it contains three expressions:
the two variables and the application itself.
This allows the model to learn that, \eg, expressions closer to the
leaves are more likely to be blamed than expressions closer to the root.

\mypara{Typing features}
A natural way of summarizing the context in which an expression occurs
is with \emph{types}.
Of course, the programs we are given are \emph{untypeable}, but we can
still extract a \emph{partial} typing derivation from the type checker
and use it to provide more information to the model.

A difficulty that arises here is that, due to the parametric type
constructors $\tfun{\cdot}{\cdot}$, $\tprod{\cdot}{\cdot}$, and
$\tlist{\cdot}$, there is an \emph{infinite} set of possible types ---
but we must have a \emph{finite} set of features.
Thus, we abstract the type of an expression to the set of type
constructors it \emph{mentions}, and add features for each type
constructor that describe whether a given type mentions the type
constructor.
For example, the type $\tint$ would only enable the $\tint$ feature,
while the type $\tfun{\tint}{\tbool}$ would enable the
$\tfun{\cdot}{\cdot}$, $\tint$, and $\tbool$ features.

We add these features for parent and child expressions to summarize the
context, but also for the current expression, as the type of an
expression is not always clear \emph{syntactically}.
For example, the expressions |tl| and |sumList tl|
in \autoref{tab:sumList} both enable \HasTypeList, as they
are both inferred to have a type that mentions $\tlist{\cdot}$.

Note that our use of typing features in an ill-typed program subjects us
to \emph{traversal bias} \citep{McAdam1998-ub}. For example, the
|sumList tl| expression might alternatively be assigned the type
$\tint$.
Our models will have to learn good localizations in spite of this bias (see
\autoref{sec:evaluation}).

\mypara{Type error slice}
Finally, we wish to distinguish between changes that could fix the
error, and changes that \emph{cannot possibly} fix the error.
Thus, we compute a minimal type error \emph{slice} for the program
(\ie the set of expressions that contribute to the error), and add a
feature that is enabled for expressions that are part of the slice.
The \InSlice feature in \autoref{tab:sumList} indicates whether an
expression is part of such a minimal slice, and is enabled for all of
the sampled expressions except for |tl|, which does not affect
the type error.
If the program contains multiple type errors, we compute
a minimal slice for each error.

In practice, we have found that \InSlice is a particularly important
feature, and thus include a post-processing step that discards all
expressions where it is disabled.
As a result, the |tl| expression would never actually be shown to the
classifier.
 We will demonstrate the importance of \InSlice empirically in
\autoref{sec:feature-utility}.

\mysubsection{Labels}
\label{sec:labels}
Recall that we make predictions in two stages.
First, we use $\evalsym$ to predict for each subexpression whether it
should be blamed, and extract a confidence score $\Runit$ from the
$\Model$.
Thus, we define the output of the $\Model$ to be a boolean label, where
``false'' means the expression \emph{should not} change and ``true''
means the expression \emph{should} change.
This allows us to predict whether any individual expression should
change, but we would actually like to predict the \emph{most likely}
expressions to change.
Second, we \emph{rank} each subexpression by the confidence $\Runit$
that it should be blamed, and return to the user the top $k$
most likely blame assignments (in practice $k=3$).


We identify the fixes for each ill-typed program with an
expression-level diff~\citep{Lempsink2009-xf}.
We consider two sources of changes.
%
First, if an expression has been removed wholesale, \eg if $\eapp{f}{x}$
  is rewritten to $\eapp{g}{x}$, we will mark the expression $f$ as
  changed, as it has been replaced by $g$.
Second, if a new expression has been inserted \emph{around} an existing
  expression, \eg if $\eapp{f}{x}$ is rewritten to
  $\eplus{\eapp{f}{x}}{1}$, we will mark the application expression
  $\eapp{f}{x}$ (but not $f$ or $x$) as changed, as the $+$ operator now
  occupies the original location of the application.

\mysubsection{Learning Algorithms}
\label{sec:models}
\lstDeleteShortInline{|} 

Recall that we formulate type error detection at a single expression as
a supervised classification problem.
This means that we are given a training data set
$S : \List{\V \times \B}$
of labeled examples, and
our goal is to use it to build a \emph{classifier}, \ie a rule
that can predict a label $b$ for an input $v$.
Since we apply the classifier on each expression in the program to
determine those that are the most likely to be type errors, we also
require the classifier to output a \emph{confidence score} that measures
how sure the classifier is about its prediction.

%
%

There are many learning algorithms to choose from, existing on a
spectrum that balances expressiveness with ease of training (and of
interpreting the learned model).
In this section we consider four standard learning algorithms: (1)
logistic regression, (2) decision trees, (3) random forests, and (4)
neural networks.
A thorough introduction to these techniques can be found in introductory
machine learning textbooks \citep[\eg][]{Hastie2009-bn}.
%

Below we briefly introduce each technique by describing the rules it
learns, and summarize its advantages and disadvantages.
For our application, we are particularly interested in three properties
-- expressiveness, interpretability and ease of generalization.
Expressiveness measures how complex prediction rules are allowed to be,
and interpretability measures how easy it is to explain the cause of
prediction to a human.
Finally ease of generalization measures how easily the rule generalizes
to examples that are not in the training set; a rule that is not
easily generalizable might perform poorly on an unseen test set even
when its training performance is high.



\mypara{Logistic Regression}
The simplest classifier we investigate is logistic regression:
a linear model where the goal is to learn a set of weights $W$
that describe the following model for predicting a label
$b$ from a feature vector $v$:
\[ \Pr(b = 1 | v) = \frac{1}{1 + e^{-W^{\top} v}} \]
The weights $W$ are learnt from training data, and the value of
$\Pr(b | v)$ naturally leads to a confidence score $\Runit$.
Logistic regression is a widely used classification algorithm, preferred
for its simplicity, ease of generalization, and interpretability.
Its main limitation is that the prediction rule is constrained to be a
linear combination of the features, and hence relatively simple.
While this can be somewhat mitigated by adding higher order (quadratic
or cubic) features, this often requires substantial domain knowledge.

\mypara{Decision Trees}
Decision tree algorithms learn a tree of binary predicates over the
features, recursively partitioning the input space until a final
classification can be made.
Each node in the tree contains a single predicate of the form
$v_j \leq t$ for some feature $v_j$ and threshold $t$, which determines
whether a given input should proceed down the left or right subtree.
Each leaf is labeled with a prediction and the fraction of
correctly-labeled training samples that would reach it; the latter
quantity can be interpreted as the decision tree's confidence in its
prediction.
This leads to a prediction rule that can be quite expressive depending
on the data used to build it.

Training a decision tree entails finding both a set of good partitioning
predicates and a good ordering of the predicates based on data.
This is usually done in a top-down greedy manner, and there are several
standard training algorithms such as C4.5 \citep{Quinlan1993-de} and
CART \citep{Breiman1984-qy}.

Another advantage of decision trees is their ease of interpretation ---
the decision rule is a white-box model that can be readily described to
a human, especially when the tree is small.
However, the main limitation is that these trees often do not generalize
well, though this can be somewhat mitigated by \emph{pruning} the tree.

%
%
%

%
%

\mypara{Random Forests}
Random forests improve generalization by training an
\emph{ensemble} of distinct decision trees and using a majority
vote to make a prediction.
The agreement among the trees forms a natural
confidence score.
Since each classifier in the ensemble is a decision tree, this still
allows for complex and expressive classifiers.

%
%
The training process involves taking $N$ random subsets of the training
data and training a separate decision tree on each subset --- the
training process for the decision trees is often modified slightly to
reduce correlation between trees, by forcing each tree to pick features
from a random subset of all features at each node.
%

The diversity of the underlying models tends to make random forests less
susceptible to the overfitting, but it
also makes the learned model more difficult to interpret.
%

\mypara{Neural Networks}
The last (and most complex) model we use is a type of neural network
called a \emph{multi-layer perceptron} (see \citet{Nielsen2015-pu} for
an introduction to neural networks).
A multi-layer perceptron can be represented as a directed acyclic
graph whose nodes are arranged in layers that are fully connected by
weighted edges.
The first layer corresponds to the input features, and the final to the
output.
%
The output of an internal node $v$ is
\[ h_v = g(\sum_{j \in N(v)} W_{jv} h_j ) \]
where $N(v)$ is the set of nodes in the previous layer that are adjacent
to $v$, $W_{jv}$ is the weight of the $(j, v)$ edge and $h_j$ is the
output of node $j$ in the previous layer.
Finally $g$ is a non-linear function, called the activation function,
which in recent work is commonly chosen to be the \emph{rectified linear
  unit} (ReLU), defined as $g(x) = \mathsf{max}(0,x)$
\citep{Nair2010-xg}.
The number of layers, the number of neurons per layer, and the
connections between layers constitute the \emph{architecture} of a
neural network.
In this work, we use relatively simple neural networks which have an
input layer, a single hidden layer and an output layer.

A major advantage of neural networks is their ability to discover
interesting combinations of features through non-linearity, which
significantly reduces the need for manual feature engineering, and
allows high expressivity.
On the other hand, this makes the networks particularly difficult to
interpret and also difficult to generalize unless vast amounts of
training data are available.

\lstMakeShortInline{|}


%% file: syntax.tex
\begin{figure}
\small
\centering
\begin{minipage}[c]{.4\textwidth}
\[
\boxed{
\begin{array}{rcl}
e & ::=    & x \spmid \efun{x}{e} \spmid \eapp{e}{e} \spmid \elet{x}{e}{e} \\
  & \spmid & n \spmid \eplus{e}{e}\\
  & \spmid & b \spmid \eif{e}{e}{e} \\
  & \spmid & \epair{e}{e} \spmid \epcase{e}{x}{x}{e} \\
  & \spmid & \enil \spmid \econs{e}{e} \spmid \ecase{e}{e}{x}{x}{e} \\[0.05in]

n & ::= &  0, 1, -1, \ldots \\[0.05in]

b & ::= &  \etrue \spmid \efalse \\[0.05in]

t & ::= & \alpha \spmid \tbool \spmid \tint \spmid \tfun{t}{t} \spmid \tprod{t}{t} \spmid \tlist{t} \\[0.05in]
\end{array}
}
\]
\captionof{figure}{Syntax of \lang}
\label{fig:syntax}
\end{minipage}
\hfill
\begin{minipage}[c]{.45\textwidth}
\lstDeleteShortInline{|} 
\[
\boxed{
\begin{array}{lcl}
  \V          & \defeq & \List{\R}\\
  \Runit      & \defeq & \{r \in \R\ |\ 0 \le r \le 1\} \\ 
  \featuresym & : & \List{e \to \R} \\
  \labelsym   & : & e \times e \to \List{e} \\
  \extractsym & : & \List{e \to \R} \to e \times e \to \List{\V \times \B} \\
  \trainsym   & : & \List{\V \times \B} \to \Model \\
  \evalsym    & : & \Model \to \V \to \Runit \\
  \midrule
  \blamesym   & : & \Model \to e \to \List{e \times \Runit}
\end{array}
}
\]
\lstMakeShortInline{|}
\captionof{figure}{
  A high-level API for converting program pairs to
  feature vectors and labels.
}
\label{fig:api}
\end{minipage}
\end{figure}

%% file: evaluation.tex
\mysection{Evaluation}
\label{sec:evaluation}
\input{data}

We have implemented our technique for localizing type errors for a
purely functional subset of \ocaml with polymorphic types and functions.
We seek to answer four questions in our evaluation:
\begin{itemize}
\item \textbf{Blame Accuracy}
  How often does \toolname
  blame a \emph{correct}
  location for the type error?
  (\autoref{sec:quantitative})
  %
\item \textbf{Feature Utility}
  Which program \emph{features are required}
  to localize errors?
   (\autoref{sec:feature-utility})
\item \textbf{Interpretability}
  %
  Are the models \emph{interpretable} using
  our intuition about type errors?
  (\autoref{sec:qualitative})

\item \textbf{Blame Utility}
  Do \toolname's blame assignments help
  users diagnose type errors?
  (\autoref{sec:user-study})
\end{itemize}
%
%
In the sequel we present our experimental
methodology \autoref{sec:methodology} and
then drill into how we evaluated each of
the questions above.
However, for the impatient reader, we begin
with a quick summary of our main results:
%
%
%

\mypara{1. Data Beats Algorithms}
Our main result is that for (novice) type error
localization, data is indeed unreasonably
effective \citep{halevy09}.
When trained on student errors from one
instance of an undergraduate course and
tested on another instance,
\toolname's most sophisticated
\emph{neural network}-based
classifier's top-ranked
prediction blames the correct
sub-term \HiddenFhTopOne\% of the time
--- a good \ToolnameWinSherrloc points
higher than the state-of-the-art
\sherrloc's \SherrlocTopOne\%.
However, even \toolname's simple
\emph{logistic regression}-based
classifier is correct \LinearTopOne\% of the time,
\ie \LinearWinSherrloc points better than \sherrloc.
When the top three predictions are considered,
\toolname is correct \HiddenFhTopThree\% of the time.

\mypara{2. Slicing Is Critical}
However, data is effective \emph{only}
when irrelevant sub-terms have been
sliced out of consideration.
In fact, perhaps our most surprising
result is that type error slicing and
local syntax alone yields
a classifier that is \SlicingWinOcaml points
better than \ocaml and on par with
\sherrloc.
That is, once we focus our classifiers on
slices, purely local syntactic features
perform as well as the
state-of-the-art.

\mypara{3. Size Doesn't Matter, Types Do}
We find that (after slices)
typing features
provide the biggest
improvement in accuracy.
Furthermore, we find contextual syntactic
features to be mostly (but not entirely)
redundant with typing features,
which supports the hypothesis that
the context's \emph{type} nicely
summarizes the properties of the
surrounding expressions.
Finally, we found that the \emph{size}
of the sub-expression was not very useful.
This was unexpected, as we thought
smaller expressions would be simpler, and
hence, more likely causes.

\mypara{4. Models Learn Typing Rules}
Finally, by investigating a few of the
predictions made by the \emph{decision tree}-based
models, we found that the models
appear to capture some simple and intuitive
rules for predicting well-typedness.
For example, if the left child of an application
is a function, then the application is likely
correct.

%

\mysubsection{Methodology}
\label{sec:methodology}

We answer our questions on two sets of data gathered from the
undergraduate Programming Languages course at
\begin{anonsuppress}
UC San Diego (IRB \#140608).
\end{anonsuppress}
\begin{noanonsuppress}
AUTHOR's INSTITUTION.
\end{noanonsuppress}
We recorded each interaction with the \ocaml top-level system while the
students worked on 23 programs from the first three homework
assignments, capturing ill-typed programs and, crucially, their
subsequent fixes.
The first dataset comes from the Spring 2014 class (\SPRING), with a
cohort of 46 students. The second comes from the Fall 2015 class
(\FALL), with a cohort of 56 students.
The extracted programs are relatively small, but they demonstrate a
range of functional programming idioms, \eg higher-order functions and
(polymorphic) algebraic data types.

\mypara{Feature Selection}
We extract 282 features from each sub-expression in a
program, including:
\begin{enumerate}
\item 45 local syntactic features. In addition to the syntax of \lang,
  we support the full range of arithmetic operators (integer and
  floating point), equality and comparison operators, character and
  string literals, and a user-defined 
  arithmetic
  expressions. We discuss the challenge of supporting other
  types in \autoref{sec:discussion}.
\item 180 contextual syntactic features. For each sub-expression we
  additionally extract the local syntactic features of its parent and
  first, second, and third (left-to-right) children. If an expression
  does not have a parent or children, these features will simply be
  disabled. If an expression has more than three children, the
  classifiers will receive no information about the additional
  children.
\item 55 typing features. In addition to the types of \lang, we support
  |int|s, |float|s, |char|s, |string|s, and the user-defined |expr|
  mentioned above. These features are extracted for each sub-expression
  and its context. 
\item One feature denoting the size of each sub-expression.
\item One feature denoting whether each sub-expression is part of the
  minimal type error slice. We use this feature as a ``hard''
  constraint, sub-expressions that are not part of the minimal slice
  will be preemptively discarded. We justify this decision in
  \autoref{sec:feature-utility}.
\end{enumerate}

\mypara{Blame Oracle}
Recall from \autoref{sec:labels} that we automatically extract a blame
oracle for each ill-typed program from the (AST) diff between it and the
student's eventual fix.
A disadvantage of using diffs in this manner is that students may have
made many, potentially unrelated, changes between compilations; at some
point the ``fix'' becomes a ``rewrite''.
We do not wish to consider the ``rewrites'' in our evaluation, so we
discard outliers where the fraction of expressions that have changed is
more than one standard deviation above the mean, establishing a diff
threshold of 40\%.
This accounts for roughly 14\% of each dataset, leaving us with
2,712 program pairs for \SPRING and 2,365 pairs for \FALL.


\mypara{Accuracy Metric}
All of the tools we compare (with the exception of the standard \ocaml
compiler) can produce a list of potential error locations.
However, in a study of fault localization techniques,
\citet{Kochhar2016-oc} show that most developers will not consider more
than around five potential error locations before falling back to manual
debugging.
Type errors are relatively simple in comparison to general fault
localization, thus we limit our evaluation to the top three predictions
of each tool.
We evaluate each tool on whether a changed expression occurred in its
top one, top two, or top three predictions.

\mypara{Blame Utility}
Finally, to test the explanatory power of our blame assigments, we
ran a user study at the University of Virginia (UVA IRB \#2014009900).
We included three problems in an exam in the Spring 2017 session of UVA's
undergraduate Programming Languages course (CS 4501).
We presented the 31 students in the course with ill-typed \ocaml\
programs and asked them to
(1) \emph{explain} the type error, and
(2) \emph{fix} the type error.
For each problem the student was given the ill-typed program and
either \sherrloc or \toolname's blame assignment, with no error message.

\input{evaluation-accuracy}
\input{evaluation-utility}
\input{threats}
\input{qualitative}

\input{evaluation-user-study}


%% file: data.tex
\pgfplotstableset{col sep=comma}

\pgfplotstableread{data/sp14-baseline.csv}{\SpringBench}
\pgfplotstablevertcat{\SpringBench}{data/sp14-ocaml-results.csv}
\pgfplotstablevertcat{\SpringBench}{data/sp14-mycroft-results.csv}
\pgfplotstablevertcat{\SpringBench}{data/sp14-sherrloc-results.csv}
\pgfplotstablevertcat{\SpringBench}{data/sp14-op+context+type+size-linear-results.csv}
\pgfplotstablevertcat{\SpringBench}{data/sp14-op+context+type+size-decision-tree-results.csv}
\pgfplotstablevertcat{\SpringBench}{data/sp14-op+context+type+size-random-forest-results.csv}
\pgfplotstablevertcat{\SpringBench}{data/sp14-op+context+type+size-hidden-10-results.csv}
\pgfplotstablevertcat{\SpringBench}{data/sp14-op+context+type+size-hidden-500-results.csv}

\pgfplotstableread{data/fa15-baseline.csv}{\FallBench}
\pgfplotstablevertcat{\FallBench}{data/fa15-ocaml-results.csv}
\pgfplotstablevertcat{\FallBench}{data/fa15-mycroft-results.csv}
\pgfplotstablevertcat{\FallBench}{data/fa15-sherrloc-results.csv}
\pgfplotstablevertcat{\FallBench}{data/fa15-op+context+type+size-linear-results.csv}
\pgfplotstablevertcat{\FallBench}{data/fa15-op+context+type+size-decision-tree-results.csv}
\pgfplotstablevertcat{\FallBench}{data/fa15-op+context+type+size-random-forest-results.csv}
\pgfplotstablevertcat{\FallBench}{data/fa15-op+context+type+size-hidden-10-results.csv}
\pgfplotstablevertcat{\FallBench}{data/fa15-op+context+type+size-hidden-500-results.csv}

\pgfplotstableread{models/linear-op+slice-no-slice.cross.csv}{\SliceLinearBench}
\pgfplotstablevertcat{\SliceLinearBench}{models/linear-op+slice.cross.csv}
\pgfplotstablevertcat{\SliceLinearBench}{models/linear-op+slice-only-slice.cross.csv}
\pgfplotstableread{models/hidden-500-op+slice-no-slice.cross.csv}{\SliceHiddenBench}
\pgfplotstablevertcat{\SliceHiddenBench}{models/hidden-500-op+slice.cross.csv}
\pgfplotstablevertcat{\SliceHiddenBench}{models/hidden-500-op+slice-only-slice.cross.csv}

\pgfplotstableread{models/linear-op.cross.csv}{\FeatureLinearBench}
\pgfplotstablevertcat{\FeatureLinearBench}{models/linear-op+size.cross.csv}
\pgfplotstablevertcat{\FeatureLinearBench}{models/linear-op+context.cross.csv}
\pgfplotstablevertcat{\FeatureLinearBench}{models/linear-op+type.cross.csv}
\pgfplotstablevertcat{\FeatureLinearBench}{models/linear-op+context+size.cross.csv}
\pgfplotstablevertcat{\FeatureLinearBench}{models/linear-op+type+size.cross.csv}
\pgfplotstablevertcat{\FeatureLinearBench}{models/linear-op+context+type.cross.csv}
\pgfplotstablevertcat{\FeatureLinearBench}{models/linear-op+context+type+size.cross.csv}
\pgfplotstableread{models/linear-op.cross.csv}{\FeatureHiddenBench}
\pgfplotstablevertcat{\FeatureHiddenBench}{models/hidden-500-op+size.cross.csv}
\pgfplotstablevertcat{\FeatureHiddenBench}{models/hidden-500-op+context.cross.csv}
\pgfplotstablevertcat{\FeatureHiddenBench}{models/hidden-500-op+type.cross.csv}
\pgfplotstablevertcat{\FeatureHiddenBench}{models/hidden-500-op+context+size.cross.csv}
\pgfplotstablevertcat{\FeatureHiddenBench}{models/hidden-500-op+type+size.cross.csv}
\pgfplotstablevertcat{\FeatureHiddenBench}{models/hidden-500-op+context+type.cross.csv}
\pgfplotstablevertcat{\FeatureHiddenBench}{models/hidden-500-op+context+type+size.cross.csv}

%% file: evaluation-accuracy.tex
\mysubsection{Blame Accuracy}
\label{sec:quantitative}

First, we compare the accuracy of our predictions to the
state of the art in type error localization.

\mypara{Baseline}
We provide two baselines for the comparison: a random choice of location
from the minimized type error slice, and the standard \ocaml compiler.

\mypara{State of the Art}
\mycroft~\citep{Loncaric2016-uk} localizes type errors by searching for
a minimal subset of typing constraints that can be removed, such that
the resulting system is satisfiable.
When multiple such subsets exist it can enumerate them, though it has no
notion of which subsets are \emph{more likely} to be correct, and thus
the order is arbitrary.
\sherrloc~\citep{Zhang2014-lv} localizes errors by searching the typing
constraint graph for constraints that participate in many unsatisfiable
paths and comparatively few satisfiable paths.
It can also enumerate multiple predictions, in descending order of
likelihood.

Comparing source locations from multiple tools with their own parsers is
not trivial.
Our experimental design gives the state of the art tools the ``benefit
of the doubt'' in two ways.
First, when evaluating \mycroft and \sherrloc, we did not consider
programs where they predicted locations that our oracle could not match
with a program expression: around 6\% of programs for \mycroft and 4\%
for \sherrloc.
Second, we similarly ignored programs where \mycroft or \sherrloc timed
out (after one minute) or where they encountered an unsupported language
feature: another 5\% for \mycroft and 12\% for \sherrloc.

\mypara{Our Classifiers}
We evaluate five classifiers, each trained on the full feature set.
%
%
These include:
%
\begin{description}
\item[\linear] A logistic regression trained with a learning rate
  $\eta = 0.001$, an $L_2$ regularization rate $\lambda = 0.001$, and a
  mini-batch size of 200.
\item[\dectree] A decision tree trained with the CART algorithm
  \citep{Breiman1984-qy} and an impurity threshold of $10^{-7}$ (used to
  avoid overfitting via early stopping).
\item[\forest] A random forest \citep{Breiman2001-wo} of 30
  estimators, with an impurity threshold of $10^{-7}$.
\item[\hiddenT and \hiddenFH] Two multi-layer perceptron neural
  networks, both trained with $\eta = 0.001$, $\lambda = 0.001$, and a
  mini-batch size of 200.
  The first MLP contains a single hidden layer of 10 neurons, and the
  second contains a hidden layer of 500 neurons.
  This gives us a measure of the complexity of the MLP's model, \ie
  if the model requires many compound features, one would expect \hiddenFH
  to outperform \hiddenT.
  %
  The neurons use rectified linear units (ReLU) as their activation
  function, a common practice in modern neural networks.
\end{description}
All classifiers were trained for 20 epochs on one dataset
--- \ie they were shown each program 20 times ---
before being evaluated on the other.
The logistic regression and MLPs were trained with the \textsc{Adam}
optimizer \citep{Kingma2014-ng}, a variant of stochastic gradient
descent that has been found to converge faster.

\input{evaluation-accuracy-graph}

\mypara{Results}
\autoref{fig:accuracy-results} shows the results of our experiment.
Localizing the type errors in our benchmarks amounted, on average, to
selecting one of 3 correct locations out of a slice of 10.
Our classifiers consistently outperform the competition, ranging from
\LinearTopOne\% Top-1 accuracy (\LinearTopThree\% Top-3)
for the \linear classifier to
\HiddenFhTopOne\% Top-1 accuracy (\HiddenFhTopThree\% Top-3)
for the \hiddenFH.\@
Our baseline of selecting at random achieves \BaselineTopOne\% Top-1
accuracy (\BaselineTopThree\% Top-3),
while \ocaml achieves a Top-1 accuracy of \OcamlTopOne\%.
Interestingly, one only needs two \emph{random} guesses to outperform
\ocaml, with \BaselineTopTwo\% accuracy.
\sherrloc outperforms both baselines, and comes close to our \linear classifier,
with \SherrlocTopOne\% Top-1 accuracy (\SherrlocTopThree\% Top 3),
while \mycroft underperforms \ocaml at \MycroftTopOne\% Top-1 accuracy.
%

Surprisingly, there is little variation in accuracy between our
classifiers.
With the exception of the \linear model, they all achieve around 70\%
Top-1 accuracy and around 90\% Top-3 accuracy.
This suggests that the model they learn is relatively simple.
In particular, notice that although the \hiddenT has $50\times$ \emph{fewer}
hidden neurons than the \hiddenFH, it only loses around 4\% accuracy.
%
We also note that our classifiers consistently perform better when
trained on the \FALL programs and tested on the \SPRING programs than
vice versa.

%% file: evaluation-accuracy-graph.tex
\definecolor{blue1}{HTML}{DEEBF7}
\definecolor{blue2}{HTML}{9ECAE1}
\definecolor{blue3}{HTML}{3182BD}
\definecolor{green1}{HTML}{E5F5E0}
\definecolor{green2}{HTML}{A1D99B}
\definecolor{green3}{HTML}{31A354}

\begin{figure}[t]
\centering
\begin{tikzpicture}
\begin{axis}[
  ybar stacked,
  width=\linewidth,
  height=7cm,
  title={Accuracy of Type Error Localization Techniques},
  ylabel={Accuracy},
  bar width=0.5cm,
  ymin=0,
  ymax=1,
  ytick={0.0, 0.1, 0.2, 0.3, 0.4, 0.5, 0.6, 0.7, 0.8, 0.9, 1.0},
  yticklabel={\pgfmathparse{\tick*100}\pgfmathprintnumber{\pgfmathresult}\,\%},
  ytick style={draw=none},
  ymajorgrids = true,
  symbolic x coords={baseline, ocaml, mycroft, sherrloc,
                     op+context+type+size/linear,
                     op+context+type+size/decision-tree,
                     op+context+type+size/random-forest,
                     op+context+type+size/hidden-10,
                     op+context+type+size/hidden-500},
  enlarge x limits=0.07,
  xtick=data,
  xtick style={draw=none},
  xticklabels={\random, \ocaml, \mycroft, \sherrloc,
               \linear, \dectree, \forest, \hiddenT, \hiddenFH},
  x tick label style={font=\small},
  y tick label style={font=\small},
  reverse legend,
  transpose legend,
  legend style={legend pos = north west, legend columns=4, font=\small},
]


\addplot[draw=black, fill=green1, bar shift=.25cm] table[x=tool, y=top-1] {\FallBench};
\addlegendentry{Top-1}
\addplot[draw=black, fill=green2, bar shift=.25cm] table[x=tool, y expr=\thisrow{top-2} - \thisrow{top-1}] {\FallBench};
\addlegendentry{Top-2}
\addplot[draw=black, fill=green3, bar shift=.25cm] table[x=tool, y expr=\thisrow{top-3} - \thisrow{top-2}] {\FallBench};
\addlegendentry{Top-3}
\addlegendimage{empty legend}
\addlegendentry{\FALL}

\resetstackedplots

\addplot[draw=black, fill=blue1, bar shift=-.25cm] table[x=tool, y=top-1] {\SpringBench};
\addlegendentry{Top-1}
\addplot[draw=black, fill=blue2, bar shift=-.25cm] table[x=tool, y expr=\thisrow{top-2} - \thisrow{top-1}] {\SpringBench};
\addlegendentry{Top-2}
\addplot[draw=black, fill=blue3, bar shift=-.25cm] table[x=tool, y expr=\thisrow{top-3} - \thisrow{top-2}] {\SpringBench};
\addlegendentry{Top-3}
\addlegendimage{empty legend}
\addlegendentry{\SPRING}

\end{axis}
\end{tikzpicture}
\caption{
  Results of our comparison of type error localization
  techniques.
  We evaluate all techniques separately on two cohorts of
  students from different instances of an undergraduate
  Programming Languages course.
  Our classifiers were trained on one cohort and evaluated on the other.
  All of our classifiers outperform the state-of-the-art techniques
  \mycroft and \sherrloc.
%
}
\label{fig:accuracy-results}
\end{figure}
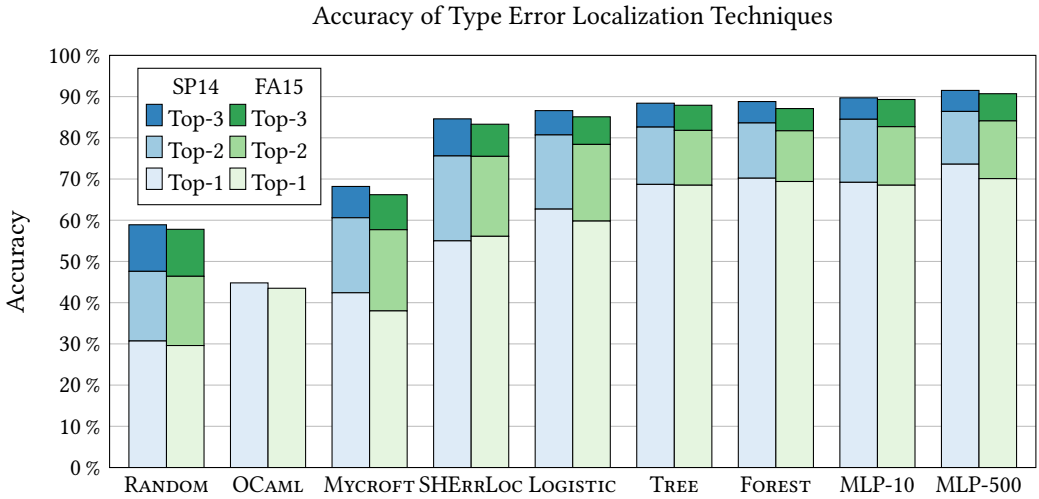

%% file: evaluation-utility.tex
\mysubsection{Feature Utility}
\label{sec:feature-utility}
We have shown that we can train a classifier to effectively localize
type errors, but which of the feature classes from
\autoref{sec:features} are contributing the most to our accuracy?
We focus specifically on feature \emph{classes} rather than individual
features as our 282 features are conceptually grouped into a much
smaller number of \emph{categorical} features.
For example, the syntactic class of an expression is conceptually a
feature but there are 45 possible values it could take; to encode this
feature for learning we split it into 45 distinct binary features.
Analyses that focus on individual features, \eg \textsc{ANOVA},
are difficult to interpret in our setting, as they will tell us the
importance of the binary features but not the higher-level categorical
features.
Thus, to answer our question we investigate the performance of
classifiers trained on various subsets of the feature classes.

\mysubsubsection{Type Error Slice}
\label{sec:type-error-slice}
First we must justify our decision to automatically exclude expressions
outside the minimal type error slice from consideration.
%
%
Thus, we compare three sets of features:
\begin{enumerate}
\item A baseline with only local syntactic features and no
  preemptive filtering by \InSlice.
\item The features of (1) extended with \InSlice.
\item The same features as (1), but we preemptively discard samples
  where \InSlice is disabled.
\end{enumerate}
The key difference between (2) and (3) is that a classifier for (2) must
\emph{learn} that \InSlice is a strong predictor.
In contrast, a classifier for (3) must only learn about the syntactic
features, the decision to discard samples where \InSlice is disabled has
already been made by a human.
This has a few additional advantages: it reduces the set of candidate
locations by a factor of 7 on average, and it guarantees that any
prediction made by the classifier can fix the type error.
We expect that (2) will perform better than (1) as it contains more
information, and that (3) will perform better than (2) as the classifier
does not have to learn the importance of \InSlice.


We tested our hypothesis with the \linear and
\hiddenFH\footnote{A layer of 500 neurons is excessive when we have so few
  input features --- we use \hiddenFH for continuity with the
  surrounding sections.}
classifiers, cross-validated ($k=10$) over the combined SP14/FA15
dataset.
%
We trained for a single epoch on feature sets (1) and (2), and for 8
epochs on (3), so that the total number of training samples would be
roughly equal for each feature set.
\lstDeleteShortInline{|} 
In addition to accuracy, we report each
classifier's \emph{recall} --- \ie ``How many true changes can we
remember?'' --- defined as
$$
\frac{|\mathsf{predicted} \cap \mathsf{oracle}|}
     {|\mathsf{oracle}|}
$$
where $\mathsf{predicted}$ is limited to the top 3 predictions, and
$\mathsf{oracle}$ is the student's fix, limited to changes that are in
the type error slice.
We make the latter distinction as:
(1) changes that are not part of the type error slice are noise in the
data set; and
(2) it makes the comparison easier to interpret since $\mathsf{oracle}$
never changes.
\lstMakeShortInline{|}
\input{evaluation-utility-graph}

\mypara{Results}
\autoref{fig:slice-utility} shows the results of our experiment.
As expected, the baseline performs the worst, with a mere 25\% \linear
Top-1 accuracy.
Adding \InSlice improves the results substantially with a 45\% \linear Top-1
accuracy, demonstrating the importance of a minimal error slice.
However, filtering out expressions that are not part of the slice
\emph{further} improves the results to 54\% \linear Top-1 accuracy.
Interestingly, while the \hiddenFH performs similarly poor with no error
slice features, it recovers nearly all of its accuracy after being given
the error slice features.
Top-1 accuracy jumps from 29\% to 53\% when we add \InSlice, and only
improves by 1\% when we filter out expressions that are not part of the
error slice.
Still, the accuracy gain comes at zero cost, and given the other benefits
of filtering by \InSlice 
--- shrinking the search space and guaranteeing our predictions are actionable ---
we choose to filter all programs by \InSlice.

\mysubsubsection{Contextual Features}
\label{sec:contextual-features}

We investigate the relative impact of the other
three classes of features discussed in \autoref{sec:features}, assuming
we have discarded expressions not in the type error slice.
For this experiment we consider again a baseline of only local syntactic
features, extended by each combination of
(1) expression size;
(2) contextual syntactic features; and
(3) typing features.
As before, we perform a 10-fold cross-validation,
but we train for a full 20 epochs to make the differences more apparent.
\mypara{Results}
\autoref{fig:context-utility} summarizes the results of this experiment.
The \linear classifier and the \hiddenFH start off
competitive when given only local syntactic features, but the \hiddenFH
quickly outperforms as we add features.

\ExprSize is the weakest feature, improving \linear Top-1
accuracy by less than 1\% and \hiddenFH by only 4\%.
In contrast, the contextual syntactic features improve \linear Top-1
accuracy by 5\% (\resp 16\%), and the typing features improve
Top-1 accuracy by 6\% (\resp 18\%).
Furthermore, while \ExprSize does provide some benefit when it is the
only additional feature, it does not appear to provide any real increase
in accuracy when added alongside the contextual or typing features.
This is likely explained by \emph{feature overlap}, \ie the contextual
features of ``child'' expressions additionally provide some information
about the size of the subtree.

As one might expect, the typing features are more beneficial than the
contextual syntactic features.
They improve Top-1 accuracy by an additional 1\% (\resp 3\%), and are much more
compact --- requiring only 55 typing features compared to 180
contextual syntactic features.
This aligns with our intuition that types should be a good summary of
the context of an expression.
However, typing features do not appear to \emph{subsume} contextual
syntactic features, the \hiddenFH gains an additional 4\% Top-1 accuracy
when both are added.

%% file: evaluation-utility-graph.tex
\begin{figure}[t]
\centering
\begin{subfigure}[t]{\linewidth}
\centering
\begin{tikzpicture}
\begin{axis}[
  name=slice,
  ybar stacked,
  width=0.5\linewidth,
  height=5cm,
  ylabel={Accuracy},
  bar width=0.5cm,
  ymin=0,
  ymax=1,
  ytick={0.0, 0.1, 0.2, 0.3, 0.4, 0.5, 0.6, 0.7, 0.8, 0.9, 1.0},
  yticklabel={\pgfmathparse{\tick*100}\pgfmathprintnumber{\pgfmathresult}\,\%},
  ytick style={draw=none},
  ymajorgrids = true,
  symbolic x coords={op+slice-no-slice, op+slice, op+slice-only-slice},
  enlarge x limits=0.25,
  xtick=data,
  xtick style={draw=none},
  xticklabels={\textsc{Local Syntax}, +\InSlice, \textsc{Filter \InSlice}},
  x tick label style={font=\small},
  y tick label style={font=\small},
  reverse legend,
  transpose legend,
  legend style={
    at={(1.75,0.5)},
    anchor=center,
    legend columns=5
  },
]


\addplot+[stack plots=false, draw=black, fill=none, thick, bar shift=.25cm] table[x=features, y=recall] {\SliceHiddenBench};
\addlegendentry{Recall}
\addplot[draw=black, fill=green1, bar shift=.25cm] table[x=features, y=top-1] {\SliceHiddenBench};
\addlegendentry{Top-1}
\addplot[draw=black, fill=green2, bar shift=.25cm] table[x=features, y expr=\thisrow{top-2} - \thisrow{top-1}] {\SliceHiddenBench};
\addlegendentry{Top-2}
\addplot[draw=black, fill=green3, bar shift=.25cm] table[x=features, y expr=\thisrow{top-3} - \thisrow{top-2}] {\SliceHiddenBench};
\addlegendentry{Top-3}
\addlegendimage{empty legend}
\addlegendentry{\hiddenFH}

\resetstackedplots

\addplot+[stack plots=false, draw=black, fill=none, thick, bar shift=-.25cm] table[x=features, y=recall] {\SliceLinearBench};
\addlegendentry{Recall}
\addplot[draw=black, fill=blue1, bar shift=-.25cm] table[x=features, y=top-1] {\SliceLinearBench};
\addlegendentry{Top-1}
\addplot[draw=black, fill=blue2, bar shift=-.25cm] table[x=features, y expr=\thisrow{top-2} - \thisrow{top-1}] {\SliceLinearBench};
\addlegendentry{Top-2}
\addplot[draw=black, fill=blue3, bar shift=-.25cm] table[x=features, y expr=\thisrow{top-3} - \thisrow{top-2}] {\SliceLinearBench};
\addlegendentry{Top-3}
\addlegendimage{empty legend}
\addlegendentry{\linear}
\end{axis}
\begin{axis}[
  ybar stacked,
  width=0.5\linewidth,
  height=5cm,
  ylabel={Recall},
  axis y line*=right,
  ymin=0,
  ymax=1,
  ytick={0.0, 0.1, 0.2, 0.3, 0.4, 0.5, 0.6, 0.7, 0.8, 0.9, 1.0},
  yticklabel={\pgfmathparse{\tick*100}\pgfmathprintnumber{\pgfmathresult}\,\%},
  ytick style={draw=none},
  ymajorgrids = false,
  xmin=0, xmax=1,
  hide x axis,
]
\end{axis}
\end{tikzpicture}
\caption{Impact of type error slice on blame accuracy.}\label{fig:slice-utility}
\end{subfigure}


\vspace{1\baselineskip}

\begin{subfigure}[t]{\linewidth}
\begin{tikzpicture}
\begin{axis}[
  ybar stacked,
  width=0.9\linewidth,
  height=5cm,
  ylabel={Accuracy},
  bar width=0.5cm,
  ymin=0.5,
  ymax=1,
  ytick={0.5, 0.6, 0.7, 0.8, 0.9, 1.0},
  yticklabel={\pgfmathparse{\tick*100}\pgfmathprintnumber{\pgfmathresult}\,\%},
  ytick style={draw=none},
  ymajorgrids = true,
  symbolic x coords={op, op+size, op+context, op+type, op+context+size, op+type+size, op+context+type, op+context+type+size},
  xtick=data,
  xtick style={draw=none},
  xticklabel style={align=center},
  xticklabels={
    \textsc{Local Syn}\\(45),
    \textsc{+Size}\\(46), \textsc{+Context}\\(225), \textsc{+Type}\\(100),
    +C+S\\(226), +T+S\\(101), +C+T\\(281),
    +C+T+S\\(282)
  },
  x tick label style={font=\small},
  y tick label style={font=\small},
]


\addplot+[stack plots=false, draw=black, fill=none, thick, bar shift=.25cm] table[x=features, y=recall] {\FeatureHiddenBench};
\addplot[draw=black, fill=green1, bar shift=.25cm] table[x=features, y=top-1] {\FeatureHiddenBench};
\addplot[draw=black, fill=green2, bar shift=.25cm] table[x=features, y expr=\thisrow{top-2} - \thisrow{top-1}] {\FeatureHiddenBench};
\addplot[draw=black, fill=green3, bar shift=.25cm] table[x=features, y expr=\thisrow{top-3} - \thisrow{top-2}] {\FeatureHiddenBench};

\resetstackedplots

\addplot+[stack plots=false, draw=black, fill=none, thick, bar shift=-.25cm] table[x=features, y=recall] {\FeatureLinearBench};
\addplot[draw=black, fill=blue1, bar shift=-.25cm] table[x=features, y=top-1] {\FeatureLinearBench};
\addplot[draw=black, fill=blue2, bar shift=-.25cm] table[x=features, y expr=\thisrow{top-2} - \thisrow{top-1}] {\FeatureLinearBench};
\addplot[draw=black, fill=blue3, bar shift=-.25cm] table[x=features, y expr=\thisrow{top-3} - \thisrow{top-2}] {\FeatureLinearBench};

\end{axis}
\begin{axis}[
  ybar stacked,
  width=0.9\linewidth,
  height=5cm,
  ylabel={Recall},
  axis y line*=right,
  ymin=0.5,
  ymax=1,
  ytick={0.5, 0.6, 0.7, 0.8, 0.9, 1.0},
  yticklabel={\pgfmathparse{\tick*100}\pgfmathprintnumber{\pgfmathresult}\,\%},
  ytick style={draw=none},
  ymajorgrids = false,
  xmin=0, xmax=1,
  hide x axis,
]
\end{axis}

\end{tikzpicture}
\caption{
  Impact of contextual features on blame accuracy.
  %
  %
  The total number of features is given in parentheses.
}
\label{fig:context-utility}
\end{subfigure}

\caption{
  Results of our experiments on feature utility.
}
\label{fig:slice-utility-results}
\end{figure}
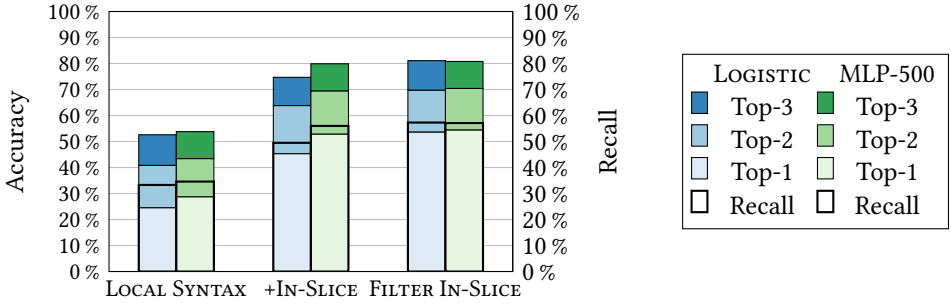
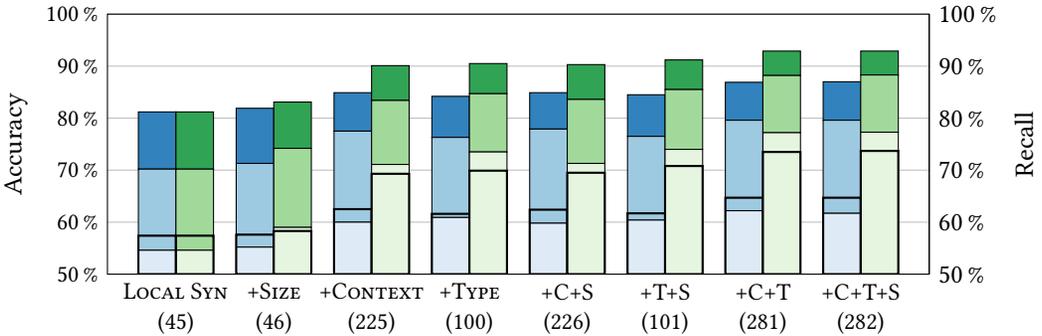

%% file: threats.tex
\mysubsection{Threats to Validity}
\label{sec:validity}

Although our experiments demonstrate that our technique can pinpoint type
errors more accurately than the state of the art and that our features are
relevant to blame assignment, our results may not generalize
to other problem domains or program sets.

One threat to validity associated with supervised machine learning is
overfitting (\ie learning a model that is too complex with respect to
the data).
A similar issue that arises in machine learning is model stability (\ie
can small changes to the training set produce large changes in the model?).
We mitigate these threats by:
(1) using separate training and testing datasets drawn from distinct
student populations (\autoref{sec:quantitative}), demonstrating the
generality of our models; and
(2) via cross-validation on the joint dataset
(\autoref{sec:feature-utility}), which demonstrates the stability of our
models by averaging the accuracy of 10 models trained on distinct
subsets of the data.

Our benchmarks were drawn from students in an undergraduate course and
may not be representative of other student populations.
We mitigate this threat by including the largest empirical evaluation of
type error localization that we are aware of: over 5,000 pairs of
ill-typed programs and fixes from two instances of the course, with
programs from 102 different students.
We acknowledge, of course, that students are not industrial programmers
and our results may not translate to large-scale software development;
however, we are particularly interested in aiding novice programmers
as they learn to work inside the type system.

A related threat to construct validity is our definition of the immedate
next well-typed program as the intended ground truth answer (see
\autoref{sec:overview}, Challenge 2). Students may, in theory, submit
intermediate well-typed program ``rewrites'' between the original ill-typed
program and the final intended answer. Our approach to discarding outliers
(see \autoref{sec:evaluation}) is designed to mitigate this threat.

Our removal of program pairs that changed too much, where our oracle
could not identify the blame of the other tools, or where the other
tools timed out or encountered unsupported language features is another
threat.
It is possible that including the programs that changed excessively
would hurt our models, or that the other tools would perform
better on the programs with unsupported language features.
We note however that
(1) outlier removal is a standard technique in machine learning%
; and
(2) our Top-1 accuracy margin is large enough that even if we assumed
that \sherrloc were perfect on all excluded programs,
we would still lead by 9 points.
%

Examining programs written in \ocaml as opposed to \haskell or any other
typed functional language poses yet another threat, common type errors
may differ in different languages.
\ocaml is, however, a standard target for research in type error
localization and thus our choice admits a direct comparison with prior
work.
Furthermore, the functional core of \ocaml that we support does not
differ significantly from the functional core of \haskell or SML, all of
which are effectively lambda calculi with a Hindley-Milner-style type
system.

Finally, our use of student fixes as oracles
assumes that students are able to correctly identify
the source of an error.
As the students are in the process of learning the language and type
system, this assumption may be faulty.
It may be that \emph{expert} users would disagree with many of the
student fixes, and that it is harder to learn a model of expert fixes,
or that the state of the art would be better at predicting expert fixes.
As we have noted before, we believe it is reasonable to use student
fixes as oracles because the student is the best judge of what she
\emph{intended}.

%% file: qualitative.tex
\mysubsection{Interpreting Specific Predictions}
\label{sec:qualitative}

Next, we present a \emph{qualitative} evaluation
that compares the predictions made by our classifiers
with those of \sherrloc.
In particular, we demonstrate, with a series of example programs from
our student dataset, how our classifiers are able to use past student
mistakes to make more accurate predictions of future fixes.
We also take this opportunity to examine some of the specific features
our classifiers use to assign blame.
For each example, we provide
(1) the code;
(2) \hlSherrloc{\sherrloc's} prediction;
(3) our \hlTree{\dectree's} prediction; and
(4) an \emph{explanation} of why our classifier made its prediction, in
terms of the features used and their values.
We choose the \dectree classifier for this section as its model is more
easily interpreted than the MLP.\@
We also exclude the \ExprSize feature from the model used in this
section, as it makes the predictions harder to motivate, and as we saw
in \autoref{sec:feature-utility} it does not appear to contribute
significantly to the model's accuracy.

We explain the predictions by analyzing the paths induced in
the decision tree by the features of the input expressions.
Recall that each node in a decision tree contains a simple predicate of
the features, \eg ``is feature $v_j$ enabled?'', which determines
whether a sample will continue down the left or right subtree.
Thus, we can examine the predicates used and the values of the
corresponding features to explain \emph{why} our \dectree made its
prediction.
We will focus particularly on the enabled features, as they generally
provide more information than the disabled features.
Furthermore, each node is additionally labeled with the ratio of
``blamed'' vs ``not-blamed'' training expressions that passed through
it.
We can use this information to identify particularly important
decisions, \ie we consider a decision that changes the ratio to be more
interesting than a decision that does not.

%

\mysubsubsection{Failed Predictions}
\label{sec:failed-predictions}

We begin with a few programs where our classifier fails to
make the correct prediction.
For these programs we will additionally \hlFix{highlight} the
correct blame location.

\mypara{Constructing a List of Duplicates}
Our first program is a simple recursive function |clone| that takes an
item |x| and a count |n|, and produces a list containing |n| copies of
|x|.
%
\begin{ecode}
  let rec clone x n =
    let loop acc n =
      if n <= 0 then
        acc
      else
        (*@\hlFix{clone} \hlSherrloc{\hlTree{([x] @\ acc)}}@*) (n - 1) in
    loop [] n
\end{ecode}
%
The student has defined a helper function |loop| with an accumulator
|acc|, likely meant to call itself tail-recursively.
Unfortunately, she has called the top-level function |clone| rather than
|loop| in the |else| branch, this induces a cyclic constraint |'a = 'a list|
for the |x| argument to |clone|.

Our top prediction coincides with \sherrloc (and \ocaml), blaming the
the first argument to |clone| rather than the occurrence of |clone| itself.
%
%
We confess that this prediction is difficult to explain by
examining the induced paths.
In particular, it only references the expression's context,
which is surprising.
More clear is why we fail to blame the occurrence of |clone|, two of
the enabled features on the path are:
(1) the parent is an application; and
(2) |clone| has a function type.
The model appears to have learned that programmers typically call the
correct function.

\mypara{Currying Considered Harmful?}
Our next example is another ill-fated attempt at |clone|.
\begin{ecode}
  let rec clone x n =
    let rec loop (*@\hlFix{x n acc}@*) =
      if n < 0 then
        acc
      else
        (*@\hlSherrloc{loop} \hlTree{(x, (n - 1), (x :: acc))}@*) in
    loop (x, n, [])
\end{ecode}
The issue here is that \ocaml functions are \emph{curried} by default
--- \ie they take their arguments one at a time --- but our student has
called the inner |loop| with all three arguments in a tuple.
Many experienced functional programmers would choose to keep |loop|
curried and rewrite the calls, however our student decides instead to
\emph{uncurry} |loop|, making it take a tuple of arguments.
\sherrloc blames the recursive call to |loop| while our classifier
blames the
tuple of arguments --- a reasonable suggestion, but not
the answer the student expected.

We fail to blame the definition of |loop| because it is defining a
function.
First, note that we represent |let f x y = e|\ \ as\ \ |let f = fun x -> fun y -> e|,
thus a change to the pattern |x| would be treated as a change to the outer
|fun| expression.
With this in mind, we can explain our failure to blame the definition of
|loop| (the outer |fun|) as follows:
(1) it has a function type;
(2) its child is a |fun|; and
(3) its parent is a |let|.
Thus it appears to the model that the outer |fun| is simply part of a
function definition, a common and innocuous phenomenon.


\mysubsubsection{Correct Predictions}
\label{sec:correct-predictions}

Next, we present a few indicative programs where our first prediction is
correct, and all of the other tools' top three predictions are
incorrect.

\mypara{Extracting the Digits of an Integer}
Consider first a simple recursive function |digitsOfInt| that extracts
the digits of an |int|.
\begin{ecode}
  let rec digitsOfInt n =
    if n <= 0 then
      []
    else
      [n mod 10] @ (*@\hlTree{[ \hlSherrloc{digitsOfInt (n / 10)} ]}@*)
\end{ecode}
Unfortunately, the student has decided to wrap the recursive call to
|digitsOfInt| with a list literal, even though |digitsOfInt| already
returns an |int list|.
Thus, the list literal is inferred to have type |int list list|, which
is incompatible with the |int list| on the left of the |@| (list append)
operator.
Both \sherrloc and the \ocaml compiler blame the recursive call for
returning a |int list| rather than |int|, but the recursive call is
correct!

As our \dectree correctly points out (with high confidence), the fault
lies with the list literal \emph{surrounding} the recursive call, remove
it and the type error disappears.
An examination of the path induced by the list literal reveals that our
\dectree is basing its decision on the fact that
(1) the expression is a list literal;
(2) the child expression is an application, whose return type mentions |int|; and
(3) the parent expression is also an application.
Interestingly, \dectree incorrectly predicts that the child application
should change as well, but it is less confident of this prediction and
ranks it below the correct blame assignment.

\mypara{Padding a list}
Our next program, |padZero|, is given two |int list|s as input, and must
left-pad the shorter one with enough zeros that the two output lists
have equal length.
The student first defines a helper |clone|.
\begin{ecode}
  let rec clone x n =
    if n <= 0 then
      []
    else
      x :: clone x (n - 1)
\end{ecode}
Then she defines |padZero| with a branch to determine which
list is shorter, followed by a |clone| to zero-pad it.
\lstset{firstnumber=last}
\begin{ecode}
  let padZero l1 l2 =
    let n = List.length l1 - List.length l2 in
    if n < 0 then
      (clone 0 ((-1) * n) @ l2, l2)
    else
      (l1, (*@\hlTree{\hlSherrloc{clone 0 n} :: l2}@*))
\end{ecode}
\lstset{firstnumber=1}
Alas, our student has accidentally used the |::| operator rather than
the |@| operator in the |else| branch.
\sherrloc and \ocaml correctly determine that she cannot cons the
|int list| returned by |clone| onto |l2|, which is another |int list|,
but they decide to blame the call to |clone|, while our
\dectree correctly blames the |::| constructor.

Examining the path induced by the |::|, we can see that our
\dectree is influenced by the fact that:
(1) |::| is a constructor; 
(2) the parent is a tuple; and
(3) the leftmost child is an application.
We note that first fact appears to be particularly significant; an
examination of the training samples that reach that decision
reveals that, before observing the \textsc{Is-Constructor} feature the
classifier is slightly in favor of predicting ``blame'', but afterwards
it is heavily in favor of predicting ``blame''.
Many of the following decisions change the balance back towards ``no
blame'' if the ``true'' path is taken, but the |::| constructor always
takes the ``false'' path.
It would appear that our \dectree has learned that constructors are
particularly suspicious, and is looking for exceptions to this 
rule.
%

Our \dectree correctly predicts that the recursive call blamed by
\sherrloc should not be blamed; a similar examination suggests that the
crucial observation is that the recursive call's parent is a data
constructor application.

%% file: evaluation-user-study.tex
\subsection{Blame Utility}
\label{sec:user-study}



We have demonstrated in the preceding sections that we can produce
more \emph{accurate} blame assignments by learning from the collective
mistakes of prior students; however, users are the final judge of
the \emph{utility} of an error message.
Thus, in this final experiment we ask whether \toolname's correct blame
assignments aid users in \emph{understanding} type errors
more than incorrect assignments.

We assigned three problems to the students in our user study: the
|padZero| and |mulByDigit| programs from \autoref{sec:qualitative}, as
well as the following |sepConcat| program
\begin{ecode}
  let rec sepConcat sep sl =
    match sl with
    | [] -> ""
    | h::t ->
        let f a x = a ^ (sep ^ x) in
        let base = (*@\hlTree{[]}@*) in
        (*@\hlSherrloc{List.fold\_left f base sl}@*)
\end{ecode}
where the student has erroneously returned the empty list, rather than
the empty string, in the base case of the fold.
%
For each problem the students were additionally given either \toolname's
correct blame assignment or \sherrloc's incorrect blame assignment,
with no error message.
%
The full user study is available in
\ifthenelse{\equal{\isTechReport}{true}}
{\autoref{sec:user-study-exams}.}
{Appendix A of the accompanying tech report~\cite{Seidel2017Learning-TechRep}.}

Due to the nature of an in-class exam, not every student answered every
question, but we always received at least 12 (out of a possible 15 or
16) responses for each problem-tool pair.
%
This session of the course
was taught in \textsc{Reason},\footnote{\url{https://reasonml.github.io}}
a dialect of \ocaml with a more C-like syntax, and thus for the study
we transcribed the programs to \textsc{Reason} syntax.

We then instructed three annotators (one of whom is an author, the others
are graduate students at UCSD) to classify the answers as
correct or incorrect.
We performed an inter-rater reliability (IRR) analysis to determine the
degree to which the annotators consistently graded the exams.
As we had more than two annotators assigning nominal (``correct'' or
``incorrect'') ratings we used Fleiss' kappa~\cite{Fleiss1971-du} to
measure IRR.\@
Fleiss' kappa is measured on a scale from $1$, indicating total
agreement, to $-1$, indicating total disagreement, with $0$ indicating
random agreement.

Finally, we used a one-sided Mann-Whitney $U$ test~\cite{Mann1947-fd} to
determine the significance of our results.
The null hypothesis was that the responses from students given
\toolname's blame were drawn from the same distribution as those
given \sherrloc's, \ie \toolname had no effect.
Since we used a one-sided test, the alternative to the null hypothesis
is that \toolname had a \emph{positive} effect on the responses.
We reject the null hypothesis in favor of the alternative if the test
produces a significance level $p < 0.05$, a standard threshold for
determining statistical significance.

\input{evaluation-user-study-graph}


\mypara{Results}
The measured kappa values were $\kappa = 0.68$ for the explanations and
$\kappa = 0.77$ for the fixes; while there is no formal notion for what
consititutes strong agreement~\cite{Krippendorff2012-wd}, kappa values
above $0.60$ are often called ``substantial''
agreement~\cite{Landis1977-ey}.
Figure~\ref{fig:results-user-study} summarizes a single annotator's
results, which show that students given \toolname's blame assignment
were generally more likely to correctly explain and fix the type error
than those given \sherrloc's.
There was no discernible difference between \toolname and
\sherrloc for |sepConcat|; however, \toolname responses for |padZero| and
|mulByDigit| were marked correct 5--25\% more often than the \sherrloc
responses.
While the results appear to show a trend in favor of \toolname,
they do not rise to the level of statistical significance
in this experiment; further investigation is merited.
%

\mypara{Threats to Validity}
Measuring understanding is difficult, and comes with its own
set of threats. 

\paragraph{Construct.}
We used the correctness of the student's explanation of, and fix for,
the type error as a proxy for her understanding, but it is possible
that other metrics would produce different results.
A further threat arises from our decision to use \textsc{Reason} syntax
rather than \ocaml.
\textsc{Reason} and \ocaml differ only in syntax, the type system is the
same; however, the difference in syntax may affect students'
understanding of the programs.
For example, \textsc{Reason} uses the notation |[h, ...t]| for the list
``cons'' constructor, in contrast to \ocaml's |h::t|.
It is quite possible that \textsc{Reason}'s syntax could help students
remember that |h| is a single element while |t| is a list.

\paragraph{Internal.}
We assigned students randomly to two groups. The first was given
\sherrloc's blame assignment for |sepConcat| and |mulByDigit|, and
\toolname's blame for |padZero|; the second was given the opposite
assignment. This ensured that each student was given \sherrloc and
\toolname problems. Students without sufficient knowledge of
\textsc{Reason} could affect the results, as could the time-constrained
nature of an exam. Thus, we excluded any answers left blank
from our analysis.

\paragraph{External.}
Our experiment used students in the process of learning \textsc{Reason},
and thus may not generalize to all developers. The three programs were
chosen manually, via a random selection and filtering of the programs
from the \SPRING dataset, where \toolname's top prediction was correct
but \sherrloc's was incorrect. A different selection of programs may
lead to different results.

\paragraph{Subjects.}
We collected exams from 31 students, though due to the nature of the
study not every student completed every problem.
The number of complete submissions was always at least 12 out of
a maximum of 15 or 16 per program-tool pair.
%

%% file: evaluation-user-study-graph.tex
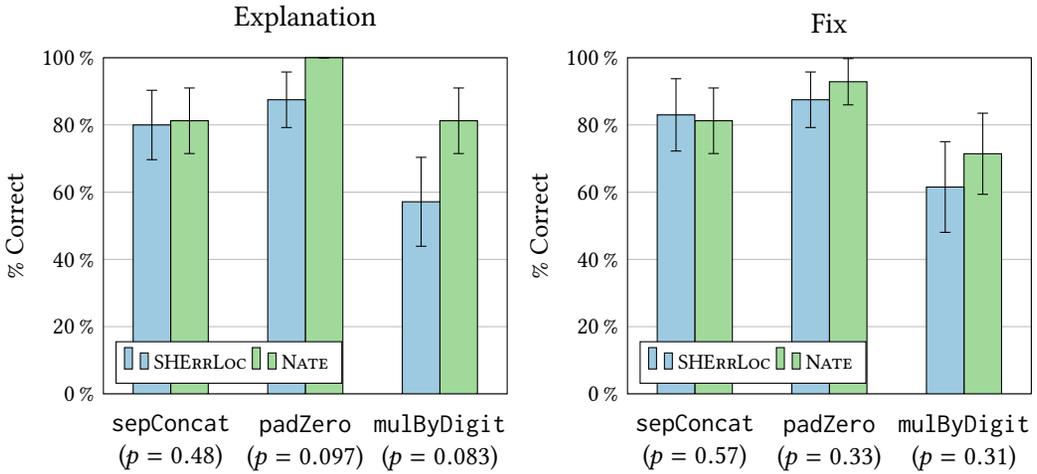
\begin{figure}[t]
\centering
\begin{tikzpicture}
\begin{axis}[
  ybar,
  height=6cm,
  title={\large Explanation},
  ylabel={\% Correct},
  bar width=0.5cm,
  ymin=0,
  ymax=1,
  yticklabel={\pgfmathparse{\tick*100}\pgfmathprintnumber{\pgfmathresult}\,\%},
  ytick style={draw=none},
  ymajorgrids = true,
  enlarge x limits=0.25,
  symbolic x coords={{sepConcat}, {padZero}, {mulByDigit}},
  xtick=data,
  xtick style={draw=none},
  xticklabel style={align=center},
  xticklabels={
    {\texttt{sepConcat}\\($p=0.48$)},
    {\texttt{padZero}\\($p=0.097$)},
    {\texttt{mulByDigit}\\($p=0.083$)}
  },
  y tick label style={font=\small},
  reverse legend,
  legend style={legend pos = south west, legend columns=2, font=\footnotesize},
]


\addplot[draw=black, fill=green2, bar shift=.25cm, error bars/.cd, y dir=both, y explicit] coordinates {
  (sepConcat, 0.8125)  +- (0, 0.0976)
  (padZero, 1.00)      +- (0, 0)
  (mulByDigit, 0.8125) +- (0, 0.0976)
};
\addlegendentry{\toolname}

\addplot[draw=black, fill=blue2, bar shift=-.25cm, error bars/.cd, y dir=both, y explicit] coordinates {
  (sepConcat, 0.80)    +- (0, 0.1033)
  (padZero, 0.875)     +- (0, 0.0827)
  (mulByDigit, 0.5714) +- (0, 0.1323)
};
\addlegendentry{\sherrloc}
\end{axis}
\end{tikzpicture}
\begin{tikzpicture}
\begin{axis}[
  ybar,
  height=6cm,
  title={\large Fix},
  ylabel={\% Correct},
  bar width=0.5cm,
  ymin=0,
  ymax=1,
  yticklabel={\pgfmathparse{\tick*100}\pgfmathprintnumber{\pgfmathresult}\,\%},
  ytick style={draw=none},
  ymajorgrids = true,
  enlarge x limits=0.25,
  symbolic x coords={{sepConcat}, {padZero}, {mulByDigit}},
  xtick=data,
  xtick style={draw=none},
  xticklabel style={align=center},
  xticklabels={
    {\texttt{sepConcat}\\($p=0.57$)},
    {\texttt{padZero}\\($p=0.33$)},
    {\texttt{mulByDigit}\\($p=0.31$)}
  },
  y tick label style={font=\small},
  reverse legend,
  legend style={legend pos = south west, legend columns=2, font=\footnotesize},
]


\addplot[draw=black, fill=green2, bar shift=.25cm, error bars/.cd, y dir=both, y explicit] coordinates {
  (sepConcat, 0.8125)  +- (0, 0.0976)
  (padZero, 0.9286)    +- (0, 0.0688)
  (mulByDigit, 0.7143) +- (0, 0.1207)
};
\addlegendentry{\toolname}

\addplot[draw=black, fill=blue2, bar shift=-.25cm, error bars/.cd, y dir=both, y explicit] coordinates {
  (sepConcat, 0.83)    +- (0, 0.1076)
  (padZero, 0.875)     +- (0, 0.0827)
  (mulByDigit, 0.6154) +- (0, 0.1349)
};
\addlegendentry{\sherrloc}
\end{axis}
\end{tikzpicture}
\caption[A classification of students' explanations and fixes for type
  errors, given either \sherrloc or \toolname's blame assignment.]
  {A classification of students' explanations and fixes for type
  errors, given either \sherrloc or \toolname's blame assignment.
  The students given \toolname's location generally scored
  better than those given \sherrloc's.
  We report the result of a one-sided Mann-Whitney $U$ test for
  statistical significance in parentheses.}
\label{fig:results-user-study}
\end{figure}

%% file: discussion.tex
\mysection{Limitations}
\label{sec:discussion}

We have shown that we can outperform the state of the art in type error
localization by learning a model of the errors that programmers make,
using a set of features that closely resemble the information the type
checker sees.
%
%
In this section we highlight some limitations of our approach and
potential avenues for future work.


\mypara{User-Defined Types}
Probably the single biggest limitation of our technique is that we have
(a finite set of) features for specific data and type constructors.
Anything our models learn about errors made with the |::| constructor or
the |list| type cannot easily be translated to new, user-defined
datatypes the model has never encountered.
We can mitigate this, to some extent, by adding generic
syntactic features for data constructors and |match| expressions, but it
remains to be seen how much these help. 
Furthermore, there is no obvious analog for transferring knowledge to
new type constructors, which we have seen are both more compact and
helpful.

As an alternative to encoding information about \emph{specific}
constructors, we might use a more abstract representation.
For example, instead of modeling |x :: 2| as a |::| constructor with a
right child of type |int|, we might model it as some (unknown) constructor
whose right child has an incompatible type.
We might symmetrically model the |2| as an integer literal whose type is
incompatible with its parent.
Anything we learn about |::| and |2| can now be transferred to
yet unseen types, but we run the risk generalizing \emph{too much} ---
\ie perhaps programmers make different errors with |list|s than they
do with other types, and are thus likely to choose different fixes.
Balancing the trade-off between specificity and generalizability appears
to be a challenging task.


\mypara{Additional Features}
There are a number of other features that could improve the model's
ability to localize errors, that would be easier to add than
user-defined types.
For example, each occurrence of a variable knows only its type and its
immediate neighbors, but it may be helpful to know
about \emph{other} occurrences of the same variable.
If a variable is generally used as a |float| but has a
single use as an |int|, it seems likely that the
latter occurrence (or context) is to blame.
Similarly, arguments to a function application are not aware of the
constraints imposed on them by the function (and vice versa),
they only know that they are occurring in the context of an application.
%
Finally, \emph{n-grams} on the token stream have proven effective for
probabilistic modeling of programming languages
\citep{Hindle2012-hf,Gabel2010-el}, we may find that they aid in
our task as well.
For example, if the observed tokens in an expression diverge from the
n-gram model's predictions, that indicates that there is something
unusual about the program at that point, and it may signal an error.


\mypara{Independent vs Joint Predictions}
We treat each sub-expression as if it exists in a vacuum, but in reality
the program has a rich \emph{graphical} structure, particularly if one adds
edges connecting different occurrences of the same variable.
\citet{Raychev2015-jg} have used these richer models to great effect to
make \emph{interdependent} predictions about programs, \eg
de-obfuscating variable names or even inferring types.
One could even view our task of locating the source of an error as simply
another property to be predicted over a graphical model of the program.
One of the key advantages of a graphical model is that the predictions
made for one node can influence the predictions made for another node,
this is known as \emph{structured learning}.
For example, if, given the expression |1 + true|, we predict |true| to
be erroneous, we may be much less likely to predict |+| as erroneous.
We compensate somewhat for our lack of structure by adding contextual
features and by ranking our predictions by ``confidence'', but it would
be interesting to see how structured learning over graphical models
would perform.


%% file: related.tex
\mysection{Related Work}
\label{sec:related-work}
\label{sec:type-error-diagnosis}

In this section we describe two relevant aspects of related work:
programming languages approaches to diagnosing type errors, and
software engineering approaches to fault localization.
%
%
%

\mypara{Localizing Type Errors}
It is well-known that the original Damas-Milner
algorithm $\mathcal{W}$ produces errors far
from their source, that novices percieve as
difficult to interpret~\citep{Wand1986-nw}.
%
%
The type checker reports an error the moment
it finds a constraint that contradicts one
of the assumptions, blaming the new inconsistent
constraint, and thus it is extremely sensitive
to the order in which it traverses the source
program (the infamous ``left-to-right''
bias~\citep{McAdam1998-ub}).
Several alternative traversal have been proposed,
\eg top-down rather than bottom-up~\citep{Lee1998-ys},
or a \emph{symmetric} traversal that checks
sub-expressions independently and only reports an
error when two inconsistent sets of constraints are
merged~\citep{McAdam1998-ub,Yang1999-yr}.
%
Type error \emph{slicing}~\citep{Haack2003-vc,Tip2001-qp,Rahli2010-ps}
overcomes the constraint-order bias by extracting a
complete and minimal subset
of terms that contribute to the error, \ie all of the
terms that are required for it to manifest and no more.
Slicing typically requires rewriting the type checker with a
specialized constraint language and solver, though
\citet{Schilling2011-yf} shows how to turn any type checker into a
slicer by treating it as a black-box.
While slicing techniques guarantee enough information to diagnose the
error, they can fall into the trap of providing \emph{too much}
information, producing a slice that is not much smaller than
the input. 

\mypara{Finding Likely Errors}
Thus, recent work has focused on finding the \emph{most likely} source
of a type error.
\citet{Zhang2014-lv} use Bayesian reasoning to search the constraint
graph for constraints that participate in many unsatisfiable paths and
relatively few satisfiable paths, based on the intuition that the
program should be mostly correct.
\citet{Pavlinovic2014-mr} translate the localization problem into a
MaxSMT problem, asking an off-the-shelf solver to find the smallest
set of constraints that can be removed such that the resulting system is
satisfiable.
\citet{Loncaric2016-uk} improve the scalability of
\citeauthor{Pavlinovic2014-mr} by reusing the existing type checker as
a theory solver in the Nelson-Oppen~\citeyear{Nelson1979-td}
style, and thus require only a MaxSAT solver.
All three of these techniques support \emph{weighted} constraints to
incorporate knowledge about the frequency of different errors,
but only \citeauthor{Pavlinovic2014-mr} use the weights, setting them to
the size of the term that induced the constraint.
In contrast, our classifiers learn a set of heuristics for predicting
the source of type errors by observing a set of ill-typed programs and
their subsequent fixes, in a sense using \emph{only} the weights and no
constraint solver.
It may be profitable to combine both approaches, \ie learn a set of good
weights for one of the above techniques from our training data.

\mypara{Explaining Type Errors}
In this paper we have focused solely on the task of \emph{localizing} a
type error, but a good error report should also \emph{explain} the
error.
\citet{Wand1986-nw}, \citet{Beaven1993-hb}, and \citet{Duggan1996-by}
attempt to explain type errors by collecting the chain of inferences
made by the type checker 
and presenting them to the user.
%
%
\citet{Gast2004-zd} produces a slice enhanced by arrows
showing the dataflow from sources with different types to a
shared sink, borrowing the insight of dataflows-as-explanations from
\textsc{MrSpidey}~\citep{Flanagan1996-bu}.
\citet{Hage2006-hc} catalog a set of heuristics for
improving the quality of error messages by examining errors made by
novices.
\citet{Heeren2003-db}, \citet{Christiansen2014-qc}, and
\citet{Serrano2016-oo} extend the ability to customize error messages to
library authors, enabling \emph{domain-specific} errors.
%
Such \emph{static} explanations of type errors run the risk of
overwhelming the user with too much information, it may be preferable to
treat type error diagnosis as an \emph{interactive} debugging session.
\citet{Bernstein1995-yj} extend the type inference procedure to handle
\emph{open} expressions (\ie with unbound variables), allowing users to
interactively query the type checker for the types of sub-expressions.
\citet{Chitil2001-td} proposes \emph{algorithmic debugging} of type
errors, presenting the user with a sequence of yes-or-no questions about
the inferred types of sub-expressions that guide the user to a specific
explanation.
\citet{Seidel2016-ul} explain type errors by searching for inputs that
expose the \emph{run-time} error that the type system prevented, and
present users with an interactive visualization of the erroneous
computation.

%

\mypara{Fixing Type Errors}
Some techniques go beyond explaining or locating a type error,
and actually attempt to \emph{fix} the error automatically.
\citet{Lerner2007-dt} searches for fixes by enumerating a
set of local mutations to the program and querying the type checker to
see if the error remains.
\citet{Chen2014-gd} use a notion of \emph{variation-based} typing to
track choices made by the type checker and enumerate potential
type (and expression)
changes that would fix the error.
They also extend the algorithmic debugging technique of
\citeauthor{Chitil2001-td} by allowing the user to enter the expected
type of specific sub-expressions and suggesting fixes based on these
desired types \citeyear{Chen2014-vm}.
Our classifiers do not attempt to suggest fixes to type errors, but it
may be possible to do so by training a classifier to predict the
syntactic class of each expression in the \emph{fixed} program --- we
believe this is an exciting direction for future work.

\mypara{Fault Localization}
Given a defect, \emph{fault localization} is the task of identifying
``suspicious'' program elements (\eg lines, statements) that are likely
implicated in the defect 
--- thus, type error localization can be viewed as an instance of fault
localization.
The best-known fault localization technique is likely Tarantula, which
uses a simple mathematical formula based on measured information from
dynamic normal and buggy runs~\citep{Jones2002-us}.
Other similar approaches, including those of \citet{Chen2002-qz} and
\citet{Abreu2006-fn,Abreu2007-mu} consider alternate features of
information or refined formulae and generally obtain more precise
results; see \citet{Wong2009-pd} for a survey.
While some researchers have approached such fault localization with an
eye toward optimality (\eg \citet{Yoo2013-rw} determine optimal
coefficients), in general such fault localization approaches are limited
by their reliance on either running tests or including relevant
features.
For example, Tarantula-based techniques require a normal and a buggy run
of the program.
By contrast, we consider incomplete programs with type errors that may
not be executed in any standard sense.
Similarly, the features available influence the classes of defects that
can be localized.
For example, a fault localization scheme based purely on control flow features
will have difficulty with cross-site scripting or SQL code injection
attacks, which follow the same control flow path on normal and buggy
runs (differing only in the user-supplied data).
Our feature set is comprised entirely of syntactic and typing features,
a natural choice for type errors, but it would likely not
generalize to other defects.



%% file: conclusion.tex
\mysection{Conclusion}
\label{sec:conclusion}

We have presented \toolname, which
combines modern statistical methods
with domain-specific feature engineering
to open the door to a new data-driven
path towards precise error localization,
significantly outperforming the
state of the art on a new benchmark
suite comprising 5,000 student programs.
%
%
We found that while machine learning
over syntactic features of each term in isolation
performs worse than existing
purely constraint-based approaches, 
augmenting the data with a single feature corresponding to
the type error slice brings our
classifiers up to par with the state of the art,
and further augmenting the data with
features of an expression's parent and children
allows our classifiers to outperform
the state of the art by \ToolnameWinSherrloc
percentage points.


As with other forms of machine learning,
a key concern is that of \emph{data-set bias}: are
\toolname's models specific
to our data set, would they fail on
\emph{other} programs?
We address this concern in two ways.
First, we partition the data by year,
and show that models learned from one
year generalize to, \ie perform nearly
as well on, the programs from the other
year.
Second, we argue that in our setting
this bias is a \emph{feature} (and not
a bug): it allows \toolname to \emph{adapt}
to the kinds of errors that programmers
(specifically novices, who are in greatest
need of precise feedback) actually make,
rather than hardwiring the biases of
experts who 
may suffer from blind spots. 
In this regard, we are particularly pleased
that our classifiers can be trained on a
modest amount of data, \ie a single course's
worth, and envision a future where each course
comes equipped with a model of its students' errors.


%% file: main.bbl

\begin{thebibliography}{00}


\ifx \showCODEN    \undefined \def \showCODEN     #1{\unskip}     \fi
\ifx \showDOI      \undefined \def \showDOI       #1{#1}\fi
\ifx \showISBNx    \undefined \def \showISBNx     #1{\unskip}     \fi
\ifx \showISBNxiii \undefined \def \showISBNxiii  #1{\unskip}     \fi
\ifx \showISSN     \undefined \def \showISSN      #1{\unskip}     \fi
\ifx \showLCCN     \undefined \def \showLCCN      #1{\unskip}     \fi
\ifx \shownote     \undefined \def \shownote      #1{#1}          \fi
\ifx \showarticletitle \undefined \def \showarticletitle #1{#1}   \fi
\ifx \showURL      \undefined \def \showURL       {\relax}        \fi
\providecommand\bibfield[2]{#2}
\providecommand\bibinfo[2]{#2}
\providecommand\natexlab[1]{#1}
\providecommand\showeprint[2][]{arXiv:#2}

\bibitem[\protect\citeauthoryear{Abreu, Zoeteweij, and van Gemund}{Abreu
  et~al\mbox{.}}{2006}]%
        {Abreu2006-fn}
\bibfield{author}{\bibinfo{person}{Rui Abreu}, \bibinfo{person}{Peter
  Zoeteweij}, {and} \bibinfo{person}{Arjan J~C van Gemund}.}
  \bibinfo{year}{2006}\natexlab{}.
\newblock \showarticletitle{An Evaluation of Similarity Coefficients for
  Software Fault Localization}. In \bibinfo{booktitle}{{\em 2006 12th Pacific
  Rim International Symposium on Dependable Computing}} {\em
  (\bibinfo{series}{PRDC '06})}. \bibinfo{pages}{39--46}.
\newblock
\showDOI{%
\url{https://doi.org/10.1109/PRDC.2006.18}}


\bibitem[\protect\citeauthoryear{Abreu, Zoeteweij, and van Gemund}{Abreu
  et~al\mbox{.}}{2007}]%
        {Abreu2007-mu}
\bibfield{author}{\bibinfo{person}{Rui Abreu}, \bibinfo{person}{Peter
  Zoeteweij}, {and} \bibinfo{person}{Arjan J~C van Gemund}.}
  \bibinfo{year}{2007}\natexlab{}.
\newblock \showarticletitle{On the Accuracy of Spectrum-based Fault
  Localization}. In \bibinfo{booktitle}{{\em Testing: Academic and Industrial
  Conference Practice and Research Techniques - {MUTATION}}} {\em
  (\bibinfo{series}{TAICPART-MUTATION 2007})}. \bibinfo{pages}{89--98}.
\newblock
\showDOI{%
\url{https://doi.org/10.1109/TAIC.PART.2007.13}}


\bibitem[\protect\citeauthoryear{Beaven and Stansifer}{Beaven and
  Stansifer}{1993}]%
        {Beaven1993-hb}
\bibfield{author}{\bibinfo{person}{Mike Beaven} {and} \bibinfo{person}{Ryan
  Stansifer}.} \bibinfo{year}{1993}\natexlab{}.
\newblock \showarticletitle{Explaining Type Errors in Polymorphic Languages}.
\newblock \bibinfo{journal}{{\em ACM Lett. Program. Lang. Syst.\/}}
  \bibinfo{volume}{2}, \bibinfo{number}{1-4} (\bibinfo{date}{March}
  \bibinfo{year}{1993}), \bibinfo{pages}{17--30}.
\newblock
\showISSN{1057-4514}
\showDOI{%
\url{https://doi.org/10.1145/176454.176460}}


\bibitem[\protect\citeauthoryear{Bernstein and Stark}{Bernstein and
  Stark}{1995}]%
        {Bernstein1995-yj}
\bibfield{author}{\bibinfo{person}{Karen~L Bernstein} {and}
  \bibinfo{person}{Eugene~W Stark}.} \bibinfo{year}{1995}\natexlab{}.
\newblock \bibinfo{booktitle}{{\em Debugging Type Errors}}.
\newblock \bibinfo{type}{{T}echnical {R}eport}. \bibinfo{institution}{State
  University of New York at Stony Brook}.
\newblock


\bibitem[\protect\citeauthoryear{Bielik, Raychev, and Vechev}{Bielik
  et~al\mbox{.}}{2016}]%
        {Bielik2016-br}
\bibfield{author}{\bibinfo{person}{Pavol Bielik}, \bibinfo{person}{Veselin
  Raychev}, {and} \bibinfo{person}{Martin Vechev}.}
  \bibinfo{year}{2016}\natexlab{}.
\newblock \showarticletitle{{PHOG}: Probabilistic Model for Code}. In
  \bibinfo{booktitle}{{\em Proceedings of the 33rd International Conference on
  Machine Learning}} {\em (\bibinfo{series}{ICML '16})}.
\newblock


\bibitem[\protect\citeauthoryear{Breiman}{Breiman}{2001}]%
        {Breiman2001-wo}
\bibfield{author}{\bibinfo{person}{Leo Breiman}.}
  \bibinfo{year}{2001}\natexlab{}.
\newblock \showarticletitle{Random Forests}.
\newblock \bibinfo{journal}{{\em Mach. Learn.\/}} \bibinfo{volume}{45},
  \bibinfo{number}{1} (\bibinfo{date}{1~Oct.} \bibinfo{year}{2001}),
  \bibinfo{pages}{5--32}.
\newblock
\showISSN{0885-6125, 1573-0565}
\showDOI{%
\url{https://doi.org/10.1023/A:1010933404324}}


\bibitem[\protect\citeauthoryear{Breiman, Friedman, Stone, and Olshen}{Breiman
  et~al\mbox{.}}{1984}]%
        {Breiman1984-qy}
\bibfield{author}{\bibinfo{person}{Leo Breiman}, \bibinfo{person}{Jerome
  Friedman}, \bibinfo{person}{Charles~J Stone}, {and}
  \bibinfo{person}{Richard~A Olshen}.} \bibinfo{year}{1984}\natexlab{}.
\newblock \bibinfo{booktitle}{{\em Classification and regression trees}}.
\newblock \bibinfo{publisher}{CRC press}.
\newblock


\bibitem[\protect\citeauthoryear{Chen, Kiciman, Fratkin, Fox, and Brewer}{Chen
  et~al\mbox{.}}{2002}]%
        {Chen2002-qz}
\bibfield{author}{\bibinfo{person}{M~Y Chen}, \bibinfo{person}{E Kiciman},
  \bibinfo{person}{E Fratkin}, \bibinfo{person}{A Fox}, {and}
  \bibinfo{person}{E Brewer}.} \bibinfo{year}{2002}\natexlab{}.
\newblock \showarticletitle{Pinpoint: problem determination in large, dynamic
  Internet services}. In \bibinfo{booktitle}{{\em Proceedings International
  Conference on Dependable Systems and Networks}}. \bibinfo{pages}{595--604}.
\newblock
\showDOI{%
\url{https://doi.org/10.1109/DSN.2002.1029005}}


\bibitem[\protect\citeauthoryear{Chen and Erwig}{Chen and Erwig}{2014a}]%
        {Chen2014-gd}
\bibfield{author}{\bibinfo{person}{Sheng Chen} {and} \bibinfo{person}{Martin
  Erwig}.} \bibinfo{year}{2014}\natexlab{a}.
\newblock \showarticletitle{Counter-factual Typing for Debugging Type Errors}.
  In \bibinfo{booktitle}{{\em Proceedings of the 41st {ACM} {SIGPLAN-SIGACT}
  Symposium on Principles of Programming Languages}} {\em
  (\bibinfo{series}{POPL '14})}. \bibinfo{publisher}{ACM},
  \bibinfo{address}{New York, NY, USA}, \bibinfo{pages}{583--594}.
\newblock
\showISBNx{9781450325448}
\showDOI{%
\url{https://doi.org/10.1145/2535838.2535863}}


\bibitem[\protect\citeauthoryear{Chen and Erwig}{Chen and Erwig}{2014b}]%
        {Chen2014-vm}
\bibfield{author}{\bibinfo{person}{Sheng Chen} {and} \bibinfo{person}{Martin
  Erwig}.} \bibinfo{year}{2014}\natexlab{b}.
\newblock \showarticletitle{Guided Type Debugging}.
\newblock In \bibinfo{booktitle}{{\em Functional and Logic Programming}},
  \bibfield{editor}{\bibinfo{person}{Michael Codish} {and}
  \bibinfo{person}{Eijiro Sumii}} (Eds.). \bibinfo{publisher}{Springer
  International Publishing}, \bibinfo{pages}{35--51}.
\newblock
\showISBNx{97833190715039783319071510, 9783319071510}
\showDOI{%
\url{https://doi.org/10.1007/978-3-319-07151-0\_3}}


\bibitem[\protect\citeauthoryear{Chitil}{Chitil}{2001}]%
        {Chitil2001-td}
\bibfield{author}{\bibinfo{person}{Olaf Chitil}.}
  \bibinfo{year}{2001}\natexlab{}.
\newblock \showarticletitle{Compositional Explanation of Types and Algorithmic
  Debugging of Type Errors}. In \bibinfo{booktitle}{{\em Proceedings of the
  Sixth {ACM} {SIGPLAN} International Conference on Functional Programming}}
  {\em (\bibinfo{series}{ICFP '01})}. \bibinfo{publisher}{ACM},
  \bibinfo{address}{New York, NY, USA}, \bibinfo{pages}{193--204}.
\newblock
\showISBNx{9781581134155}
\showDOI{%
\url{https://doi.org/10.1145/507635.507659}}


\bibitem[\protect\citeauthoryear{Christiansen}{Christiansen}{2014}]%
        {Christiansen2014-qc}
\bibfield{author}{\bibinfo{person}{David~Raymond Christiansen}.}
  \bibinfo{year}{2014}\natexlab{}.
\newblock \showarticletitle{Reflect on your mistakes! Lightweight
  domain-specific error messages}. In \bibinfo{booktitle}{{\em Preproceedings
  of the 15th Symposium on Trends in Functional Programming}}.
\newblock


\bibitem[\protect\citeauthoryear{Cousot and Cousot}{Cousot and Cousot}{1977}]%
        {CousotCousot77}
\bibfield{author}{\bibinfo{person}{P. Cousot} {and} \bibinfo{person}{R.
  Cousot}.} \bibinfo{year}{1977}\natexlab{}.
\newblock \showarticletitle{Abstract interpretation: a unified lattice model
  for the static analysis of programs}. In \bibinfo{booktitle}{{\em POPL 77}}.
  \bibinfo{publisher}{ACM}, \bibinfo{pages}{238--252}.
\newblock


\bibitem[\protect\citeauthoryear{Duggan and Bent}{Duggan and Bent}{1996}]%
        {Duggan1996-by}
\bibfield{author}{\bibinfo{person}{Dominic Duggan} {and}
  \bibinfo{person}{Frederick Bent}.} \bibinfo{year}{1996}\natexlab{}.
\newblock \showarticletitle{Explaining type inference}.
\newblock \bibinfo{journal}{{\em Science of Computer Programming\/}}
  \bibinfo{volume}{27}, \bibinfo{number}{1} (\bibinfo{date}{July}
  \bibinfo{year}{1996}), \bibinfo{pages}{37--83}.
\newblock
\showISSN{0167-6423}
\showDOI{%
\url{https://doi.org/10.1016/0167-6423(95)00007-0}}


\bibitem[\protect\citeauthoryear{Flanagan, Flatt, Krishnamurthi, Weirich, and
  Felleisen}{Flanagan et~al\mbox{.}}{1996}]%
        {Flanagan1996-bu}
\bibfield{author}{\bibinfo{person}{Cormac Flanagan}, \bibinfo{person}{Matthew
  Flatt}, \bibinfo{person}{Shriram Krishnamurthi}, \bibinfo{person}{Stephanie
  Weirich}, {and} \bibinfo{person}{Matthias Felleisen}.}
  \bibinfo{year}{1996}\natexlab{}.
\newblock \showarticletitle{Catching bugs in the web of program invariants}. In
  \bibinfo{booktitle}{{\em Proceedings of the {ACM} {SIGPLAN} 1996 conference
  on Programming language design and implementation}} {\em
  (\bibinfo{series}{PLDI '96})}, Vol.~\bibinfo{volume}{31}.
  \bibinfo{publisher}{ACM}, \bibinfo{pages}{23--32}.
\newblock
\showISBNx{9780897917957}
\showISSN{0362-1340}
\showDOI{%
\url{https://doi.org/10.1145/249069.231387}}


\bibitem[\protect\citeauthoryear{Fleiss}{Fleiss}{1971}]%
        {Fleiss1971-du}
\bibfield{author}{\bibinfo{person}{Joseph~L Fleiss}.}
  \bibinfo{year}{1971}\natexlab{}.
\newblock \showarticletitle{Measuring nominal scale agreement among many
  raters}.
\newblock \bibinfo{journal}{{\em Psychol. Bull.\/}} \bibinfo{volume}{76},
  \bibinfo{number}{5} (\bibinfo{date}{Nov.} \bibinfo{year}{1971}),
  \bibinfo{pages}{378}.
\newblock
\showISSN{0033-2909, 1939-1455}
\showDOI{%
\url{https://doi.org/10.1037/h0031619}}


\bibitem[\protect\citeauthoryear{Gabel and Su}{Gabel and Su}{2010}]%
        {Gabel2010-el}
\bibfield{author}{\bibinfo{person}{Mark Gabel} {and} \bibinfo{person}{Zhendong
  Su}.} \bibinfo{year}{2010}\natexlab{}.
\newblock \showarticletitle{A Study of the Uniqueness of Source Code}. In
  \bibinfo{booktitle}{{\em Proceedings of the Eighteenth {ACM} {SIGSOFT}
  International Symposium on Foundations of Software Engineering}} {\em
  (\bibinfo{series}{FSE '10})}. \bibinfo{publisher}{ACM}, \bibinfo{address}{New
  York, NY, USA}, \bibinfo{pages}{147--156}.
\newblock
\showISBNx{9781605587912}
\showDOI{%
\url{https://doi.org/10.1145/1882291.1882315}}


\bibitem[\protect\citeauthoryear{Gast}{Gast}{2004}]%
        {Gast2004-zd}
\bibfield{author}{\bibinfo{person}{Holger Gast}.}
  \bibinfo{year}{2004}\natexlab{}.
\newblock \showarticletitle{Explaining {ML} Type Errors by Data Flows}.
\newblock In \bibinfo{booktitle}{{\em Implementation and Application of
  Functional Languages}}. \bibinfo{publisher}{Springer Berlin Heidelberg},
  \bibinfo{pages}{72--89}.
\newblock
\showISBNx{9783540260943, 9783540320388}
\showDOI{%
\url{https://doi.org/10.1007/11431664\_5}}


\bibitem[\protect\citeauthoryear{Haack and Wells}{Haack and Wells}{2003}]%
        {Haack2003-vc}
\bibfield{author}{\bibinfo{person}{Christian Haack} {and} \bibinfo{person}{J~B
  Wells}.} \bibinfo{year}{2003}\natexlab{}.
\newblock \showarticletitle{Type Error Slicing in Implicitly Typed
  {Higher-Order} Languages}.
\newblock In \bibinfo{booktitle}{{\em Programming Languages and Systems}}.
  \bibinfo{publisher}{Springer Berlin Heidelberg}, \bibinfo{pages}{284--301}.
\newblock
\showISBNx{9783540008866, 9783540365754}
\showDOI{%
\url{https://doi.org/10.1007/3-540-36575-3\_20}}


\bibitem[\protect\citeauthoryear{Hage and Heeren}{Hage and Heeren}{2006}]%
        {Hage2006-hc}
\bibfield{author}{\bibinfo{person}{Jurriaan Hage} {and}
  \bibinfo{person}{Bastiaan Heeren}.} \bibinfo{year}{2006}\natexlab{}.
\newblock \showarticletitle{Heuristics for Type Error Discovery and Recovery}.
\newblock In \bibinfo{booktitle}{{\em Implementation and Application of
  Functional Languages}}. \bibinfo{publisher}{Springer Berlin Heidelberg},
  \bibinfo{pages}{199--216}.
\newblock
\showISBNx{9783540741299, 9783540741305}
\showDOI{%
\url{https://doi.org/10.1007/978-3-540-74130-5\_12}}


\bibitem[\protect\citeauthoryear{Halevy, Norvig, and Pereira}{Halevy
  et~al\mbox{.}}{2009}]%
        {halevy09}
\bibfield{author}{\bibinfo{person}{Alon Halevy}, \bibinfo{person}{Peter
  Norvig}, {and} \bibinfo{person}{Fernando Pereira}.}
  \bibinfo{year}{2009}\natexlab{}.
\newblock \showarticletitle{The unreasonable effectiveness of data}.
\newblock \bibinfo{journal}{{\em IEEE Intelligent Systems\/}}
  \bibinfo{volume}{24}, \bibinfo{number}{2} (\bibinfo{year}{2009}),
  \bibinfo{pages}{8--12}.
\newblock


\bibitem[\protect\citeauthoryear{Hastie, Tibshirani, and Friedman}{Hastie
  et~al\mbox{.}}{2009}]%
        {Hastie2009-bn}
\bibfield{author}{\bibinfo{person}{Trevor Hastie}, \bibinfo{person}{Robert
  Tibshirani}, {and} \bibinfo{person}{Jerome Friedman}.}
  \bibinfo{year}{2009}\natexlab{}.
\newblock \bibinfo{booktitle}{{\em The Elements of Statistical Learning: Data
  Mining, Inference, and Prediction}}.
\newblock \bibinfo{publisher}{Springer New York}.
\newblock
\showISBNx{9780387848570, 9780387848587}
\showDOI{%
\url{https://doi.org/10.1007/978-0-387-84858-7}}


\bibitem[\protect\citeauthoryear{Heeren, Hage, and Swierstra}{Heeren
  et~al\mbox{.}}{2003}]%
        {Heeren2003-db}
\bibfield{author}{\bibinfo{person}{Bastiaan Heeren}, \bibinfo{person}{Jurriaan
  Hage}, {and} \bibinfo{person}{S~Doaitse Swierstra}.}
  \bibinfo{year}{2003}\natexlab{}.
\newblock \showarticletitle{Scripting the type inference process}. In
  \bibinfo{booktitle}{{\em Proceedings of the eighth {ACM} {SIGPLAN}
  international conference on Functional programming}},
  Vol.~\bibinfo{volume}{38}. \bibinfo{publisher}{ACM}, \bibinfo{pages}{3--13}.
\newblock
\showISBNx{9781581137569}
\showISSN{0362-1340}
\showDOI{%
\url{https://doi.org/10.1145/944705.944707}}


\bibitem[\protect\citeauthoryear{Hindle, Barr, Su, Gabel, and Devanbu}{Hindle
  et~al\mbox{.}}{2012a}]%
        {Devanbu:2012}
\bibfield{author}{\bibinfo{person}{Abram Hindle}, \bibinfo{person}{Earl~T.
  Barr}, \bibinfo{person}{Zhendong Su}, \bibinfo{person}{Mark Gabel}, {and}
  \bibinfo{person}{Premkumar Devanbu}.} \bibinfo{year}{2012}\natexlab{a}.
\newblock \showarticletitle{On the Naturalness of Software}. In
  \bibinfo{booktitle}{{\em Proceedings of the 34th International Conference on
  Software Engineering}} {\em (\bibinfo{series}{ICSE '12})}.
  \bibinfo{publisher}{IEEE Press}, \bibinfo{address}{Piscataway, NJ, USA},
  \bibinfo{pages}{837--847}.
\newblock
\showISBNx{978-1-4673-1067-3}
\showURL{%
\url{http://dl.acm.org/citation.cfm?id=2337223.2337322}}


\bibitem[\protect\citeauthoryear{Hindle, Barr, Su, Gabel, and Devanbu}{Hindle
  et~al\mbox{.}}{2012b}]%
        {Hindle2012-hf}
\bibfield{author}{\bibinfo{person}{Abram Hindle}, \bibinfo{person}{Earl~T
  Barr}, \bibinfo{person}{Zhendong Su}, \bibinfo{person}{Mark Gabel}, {and}
  \bibinfo{person}{Premkumar Devanbu}.} \bibinfo{year}{2012}\natexlab{b}.
\newblock \showarticletitle{On the Naturalness of Software}. In
  \bibinfo{booktitle}{{\em Proceedings of the 34th International Conference on
  Software Engineering}} {\em (\bibinfo{series}{ICSE '12})}.
  \bibinfo{publisher}{IEEE Press}, \bibinfo{address}{Piscataway, NJ, USA},
  \bibinfo{pages}{837--847}.
\newblock
\showISBNx{9781467310673}


\bibitem[\protect\citeauthoryear{Jones, Harrold, and Stasko}{Jones
  et~al\mbox{.}}{2002}]%
        {Jones2002-us}
\bibfield{author}{\bibinfo{person}{James~A Jones}, \bibinfo{person}{Mary~Jean
  Harrold}, {and} \bibinfo{person}{John Stasko}.}
  \bibinfo{year}{2002}\natexlab{}.
\newblock \showarticletitle{Visualization of Test Information to Assist Fault
  Localization}. In \bibinfo{booktitle}{{\em Proceedings of the 24th
  International Conference on Software Engineering}} {\em
  (\bibinfo{series}{ICSE '02})}. \bibinfo{publisher}{ACM},
  \bibinfo{address}{New York, NY, USA}, \bibinfo{pages}{467--477}.
\newblock
\showISBNx{9781581134728}
\showDOI{%
\url{https://doi.org/10.1145/581339.581397}}


\bibitem[\protect\citeauthoryear{Joosten, Van Den~Berg, and Van
  Der~Hoeven}{Joosten et~al\mbox{.}}{1993}]%
        {Joosten1993-yx}
\bibfield{author}{\bibinfo{person}{Stef Joosten}, \bibinfo{person}{Klaas Van
  Den~Berg}, {and} \bibinfo{person}{Gerrit Van Der~Hoeven}.}
  \bibinfo{year}{1993}\natexlab{}.
\newblock \showarticletitle{Teaching functional programming to first-year
  students}.
\newblock \bibinfo{journal}{{\em J. Funct. Programming\/}} \bibinfo{volume}{3},
  \bibinfo{number}{01} (\bibinfo{date}{Jan.} \bibinfo{year}{1993}),
  \bibinfo{pages}{49--65}.
\newblock
\showISSN{0956-7968, 1469-7653}
\showDOI{%
\url{https://doi.org/10.1017/S0956796800000599}}


\bibitem[\protect\citeauthoryear{Jose and Majumdar}{Jose and Majumdar}{2011}]%
        {Jose:2011}
\bibfield{author}{\bibinfo{person}{Manu Jose} {and} \bibinfo{person}{Rupak
  Majumdar}.} \bibinfo{year}{2011}\natexlab{}.
\newblock \showarticletitle{Cause Clue Clauses: Error Localization Using
  Maximum Satisfiability}.
\newblock \bibinfo{journal}{{\em SIGPLAN Not.\/}} \bibinfo{volume}{46},
  \bibinfo{number}{6} (\bibinfo{date}{June} \bibinfo{year}{2011}),
  \bibinfo{pages}{437--446}.
\newblock
\showISSN{0362-1340}
\showDOI{%
\url{https://doi.org/10.1145/1993316.1993550}}


\bibitem[\protect\citeauthoryear{Kingma and Ba}{Kingma and Ba}{2014}]%
        {Kingma2014-ng}
\bibfield{author}{\bibinfo{person}{Diederik~P Kingma} {and}
  \bibinfo{person}{Jimmy Ba}.} \bibinfo{year}{2014}\natexlab{}.
\newblock \showarticletitle{Adam: A Method for Stochastic Optimization}.
\newblock  (\bibinfo{date}{22~Dec.} \bibinfo{year}{2014}).
\newblock
\showeprint[arxiv]{1412.6980}


\bibitem[\protect\citeauthoryear{Kochhar, Xia, Lo, and Li}{Kochhar
  et~al\mbox{.}}{2016}]%
        {Kochhar2016-oc}
\bibfield{author}{\bibinfo{person}{Pavneet~Singh Kochhar}, \bibinfo{person}{Xin
  Xia}, \bibinfo{person}{David Lo}, {and} \bibinfo{person}{Shanping Li}.}
  \bibinfo{year}{2016}\natexlab{}.
\newblock \showarticletitle{Practitioners' Expectations on Automated Fault
  Localization}. In \bibinfo{booktitle}{{\em Proceedings of the 25th
  International Symposium on Software Testing and Analysis}} {\em
  (\bibinfo{series}{ISSTA 2016})}. \bibinfo{publisher}{ACM},
  \bibinfo{address}{New York, NY, USA}, \bibinfo{pages}{165--176}.
\newblock
\showISBNx{9781450343909}
\showDOI{%
\url{https://doi.org/10.1145/2931037.2931051}}


\bibitem[\protect\citeauthoryear{Kotsiantis}{Kotsiantis}{2007}]%
        {Kotsiantis2007-pj}
\bibfield{author}{\bibinfo{person}{S~B Kotsiantis}.}
  \bibinfo{year}{2007}\natexlab{}.
\newblock \showarticletitle{Supervised Machine Learning: A Review of
  Classification Techniques}.
\newblock \bibinfo{journal}{{\em Informatica\/}} \bibinfo{volume}{31},
  \bibinfo{number}{3} (\bibinfo{year}{2007}), \bibinfo{pages}{249--268}.
\newblock
\showISSN{0350-5596}


\bibitem[\protect\citeauthoryear{Kremenek and Engler}{Kremenek and
  Engler}{2003}]%
        {Kremenek2003-ck}
\bibfield{author}{\bibinfo{person}{Ted Kremenek} {and} \bibinfo{person}{Dawson
  Engler}.} \bibinfo{year}{2003}\natexlab{}.
\newblock \showarticletitle{{Z-Ranking}: Using Statistical Analysis to Counter
  the Impact of Static Analysis Approximations}.
\newblock In \bibinfo{booktitle}{{\em Static Analysis}},
  \bibfield{editor}{\bibinfo{person}{Radhia Cousot}} (Ed.).
  \bibinfo{series}{Lecture Notes in Computer Science},
  Vol.~\bibinfo{volume}{2694}. \bibinfo{publisher}{Springer Berlin Heidelberg},
  \bibinfo{address}{Berlin, Heidelberg}, \bibinfo{pages}{295--315}.
\newblock
\showISBNx{9783540403258, 9783540448983}
\showISSN{0302-9743}
\showDOI{%
\url{https://doi.org/10.1007/3-540-44898-5\_16}}


\bibitem[\protect\citeauthoryear{Krippendorff}{Krippendorff}{2012}]%
        {Krippendorff2012-wd}
\bibfield{author}{\bibinfo{person}{K Krippendorff}.}
  \bibinfo{year}{2012}\natexlab{}.
\newblock \bibinfo{booktitle}{{\em Content Analysis: An Introduction to Its
  Methodology}}.
\newblock \bibinfo{publisher}{SAGE Publications}.
\newblock
\showISBNx{9781412983150}
\showLCCN{2011048278}


\bibitem[\protect\citeauthoryear{Landis and Koch}{Landis and Koch}{1977}]%
        {Landis1977-ey}
\bibfield{author}{\bibinfo{person}{J~R Landis} {and} \bibinfo{person}{G~G
  Koch}.} \bibinfo{year}{1977}\natexlab{}.
\newblock \showarticletitle{The measurement of observer agreement for
  categorical data}.
\newblock \bibinfo{journal}{{\em Biometrics\/}} \bibinfo{volume}{33},
  \bibinfo{number}{1} (\bibinfo{date}{March} \bibinfo{year}{1977}),
  \bibinfo{pages}{159--174}.
\newblock
\showISSN{0006-341X}


\bibitem[\protect\citeauthoryear{Lee and Yi}{Lee and Yi}{1998}]%
        {Lee1998-ys}
\bibfield{author}{\bibinfo{person}{Oukseh Lee} {and} \bibinfo{person}{Kwangkeun
  Yi}.} \bibinfo{year}{1998}\natexlab{}.
\newblock \showarticletitle{Proofs About a Folklore Let-polymorphic Type
  Inference Algorithm}.
\newblock \bibinfo{journal}{{\em ACM Trans. Program. Lang. Syst.\/}}
  \bibinfo{volume}{20}, \bibinfo{number}{4} (\bibinfo{date}{July}
  \bibinfo{year}{1998}), \bibinfo{pages}{707--723}.
\newblock
\showISSN{0164-0925}
\showDOI{%
\url{https://doi.org/10.1145/291891.291892}}


\bibitem[\protect\citeauthoryear{Lempsink}{Lempsink}{2009}]%
        {Lempsink2009-xf}
\bibfield{author}{\bibinfo{person}{Eelco Lempsink}.}
  \bibinfo{year}{2009}\natexlab{}.
\newblock {\em \bibinfo{title}{Generic type-safe diff and patch for families of
  datatypes}}.
\newblock \bibinfo{thesistype}{Master's\ thesis}. \bibinfo{school}{Universiteit
  Utrecht}.
\newblock


\bibitem[\protect\citeauthoryear{Lerner, Flower, Grossman, and Chambers}{Lerner
  et~al\mbox{.}}{2007}]%
        {Lerner2007-dt}
\bibfield{author}{\bibinfo{person}{Benjamin~S Lerner}, \bibinfo{person}{Matthew
  Flower}, \bibinfo{person}{Dan Grossman}, {and} \bibinfo{person}{Craig
  Chambers}.} \bibinfo{year}{2007}\natexlab{}.
\newblock \showarticletitle{Searching for Type-error Messages}. In
  \bibinfo{booktitle}{{\em Proceedings of the 28th {ACM} {SIGPLAN} Conference
  on Programming Language Design and Implementation}} {\em
  (\bibinfo{series}{PLDI '07})}. \bibinfo{publisher}{ACM},
  \bibinfo{address}{New York, NY, USA}, \bibinfo{pages}{425--434}.
\newblock
\showISBNx{9781595936332}
\showDOI{%
\url{https://doi.org/10.1145/1250734.1250783}}


\bibitem[\protect\citeauthoryear{Loncaric, Chandra, Schlesinger, and
  Sridharan}{Loncaric et~al\mbox{.}}{2016}]%
        {Loncaric2016-uk}
\bibfield{author}{\bibinfo{person}{Calvin Loncaric}, \bibinfo{person}{Satish
  Chandra}, \bibinfo{person}{Cole Schlesinger}, {and} \bibinfo{person}{Manu
  Sridharan}.} \bibinfo{year}{2016}\natexlab{}.
\newblock \showarticletitle{A practical framework for type inference error
  explanation}. In \bibinfo{booktitle}{{\em Proceedings of the 2016 {ACM}
  {SIGPLAN} International Conference on {Object-Oriented} Programming, Systems,
  Languages, and Applications}}. \bibinfo{publisher}{ACM},
  \bibinfo{pages}{781--799}.
\newblock
\showISBNx{9781450344449}
\showDOI{%
\url{https://doi.org/10.1145/2983990.2983994}}


\bibitem[\protect\citeauthoryear{Mann and Whitney}{Mann and Whitney}{1947}]%
        {Mann1947-fd}
\bibfield{author}{\bibinfo{person}{H~B Mann} {and} \bibinfo{person}{D~R
  Whitney}.} \bibinfo{year}{1947}\natexlab{}.
\newblock \showarticletitle{On a Test of Whether one of Two Random Variables is
  Stochastically Larger than the Other}.
\newblock \bibinfo{journal}{{\em Ann. Math. Stat.\/}} \bibinfo{volume}{18},
  \bibinfo{number}{1} (\bibinfo{date}{March} \bibinfo{year}{1947}),
  \bibinfo{pages}{50--60}.
\newblock
\showISSN{0003-4851, 2168-8990}
\showDOI{%
\url{https://doi.org/10.1214/aoms/1177730491}}


\bibitem[\protect\citeauthoryear{McAdam}{McAdam}{1998}]%
        {McAdam1998-ub}
\bibfield{author}{\bibinfo{person}{Bruce~J McAdam}.}
  \bibinfo{year}{1998}\natexlab{}.
\newblock \showarticletitle{On the Unification of Substitutions in Type
  Inference}. In \bibinfo{booktitle}{{\em Implementation of Functional
  Languages}} {\em (\bibinfo{series}{Lecture Notes in Computer Science})},
  \bibfield{editor}{\bibinfo{person}{Kevin Hammond}, \bibinfo{person}{Tony
  Davie}, {and} \bibinfo{person}{Chris Clack}} (Eds.).
  \bibinfo{publisher}{Springer Berlin Heidelberg}, \bibinfo{pages}{137--152}.
\newblock
\showISBNx{9783540662297, 9783540485155}
\showDOI{%
\url{https://doi.org/10.1007/3-540-48515-5\_9}}


\bibitem[\protect\citeauthoryear{Nair and Hinton}{Nair and Hinton}{2010}]%
        {Nair2010-xg}
\bibfield{author}{\bibinfo{person}{Vinod Nair} {and}
  \bibinfo{person}{Geoffrey~E Hinton}.} \bibinfo{year}{2010}\natexlab{}.
\newblock \showarticletitle{Rectified linear units improve restricted boltzmann
  machines}. In \bibinfo{booktitle}{{\em Proceedings of the 27th international
  conference on machine learning ({ICML-10})}}. \bibinfo{pages}{807--814}.
\newblock


\bibitem[\protect\citeauthoryear{Nelson and Oppen}{Nelson and Oppen}{1979}]%
        {Nelson1979-td}
\bibfield{author}{\bibinfo{person}{Greg Nelson} {and} \bibinfo{person}{Derek~C
  Oppen}.} \bibinfo{year}{1979}\natexlab{}.
\newblock \showarticletitle{Simplification by Cooperating Decision Procedures}.
\newblock \bibinfo{journal}{{\em ACM Trans. Program. Lang. Syst.\/}}
  \bibinfo{volume}{1}, \bibinfo{number}{2} (\bibinfo{date}{Oct.}
  \bibinfo{year}{1979}), \bibinfo{pages}{245--257}.
\newblock
\showISSN{0164-0925}
\showDOI{%
\url{https://doi.org/10.1145/357073.357079}}


\bibitem[\protect\citeauthoryear{Nielsen}{Nielsen}{2015}]%
        {Nielsen2015-pu}
\bibfield{author}{\bibinfo{person}{Michael~A Nielsen}.}
  \bibinfo{year}{2015}\natexlab{}.
\newblock \bibinfo{booktitle}{{\em Neural Networks and Deep Learning}}.
\newblock \bibinfo{publisher}{Determination Press}.
\newblock


\bibitem[\protect\citeauthoryear{Pavlinovic, King, and Wies}{Pavlinovic
  et~al\mbox{.}}{2014}]%
        {Pavlinovic2014-mr}
\bibfield{author}{\bibinfo{person}{Zvonimir Pavlinovic}, \bibinfo{person}{Tim
  King}, {and} \bibinfo{person}{Thomas Wies}.} \bibinfo{year}{2014}\natexlab{}.
\newblock \showarticletitle{Finding Minimum Type Error Sources}. In
  \bibinfo{booktitle}{{\em Proceedings of the 2014 {ACM} International
  Conference on Object Oriented Programming Systems Languages \& Applications}}
  {\em (\bibinfo{series}{OOPSLA '14})}. \bibinfo{publisher}{ACM},
  \bibinfo{address}{New York, NY, USA}, \bibinfo{pages}{525--542}.
\newblock
\showISBNx{9781450325851}
\showDOI{%
\url{https://doi.org/10.1145/2660193.2660230}}


\bibitem[\protect\citeauthoryear{Quinlan}{Quinlan}{1993}]%
        {Quinlan1993-de}
\bibfield{author}{\bibinfo{person}{John~Ross Quinlan}.}
  \bibinfo{year}{1993}\natexlab{}.
\newblock \bibinfo{booktitle}{{\em C4.5: Programs for Machine Learning}}.
\newblock \bibinfo{publisher}{Morgan Kaufmann}.
\newblock
\showISBNx{9781558602380}


\bibitem[\protect\citeauthoryear{Rahli, Wells, and Kamareddine}{Rahli
  et~al\mbox{.}}{2010}]%
        {Rahli2010-ps}
\bibfield{author}{\bibinfo{person}{Vincent Rahli}, \bibinfo{person}{J~B Wells},
  {and} \bibinfo{person}{Fairouz Kamareddine}.}
  \bibinfo{year}{2010}\natexlab{}.
\newblock \bibinfo{booktitle}{{\em A constraint system for a {SML} type error
  slicer}}.
\newblock \bibinfo{type}{{T}echnical {R}eport} HW-MACS-TR-0079.
  \bibinfo{institution}{Herriot Watt University}.
\newblock


\bibitem[\protect\citeauthoryear{Raychev, Vechev, and Krause}{Raychev
  et~al\mbox{.}}{2015}]%
        {Raychev2015-jg}
\bibfield{author}{\bibinfo{person}{Veselin Raychev}, \bibinfo{person}{Martin
  Vechev}, {and} \bibinfo{person}{Andreas Krause}.}
  \bibinfo{year}{2015}\natexlab{}.
\newblock \showarticletitle{Predicting Program Properties from ``Big Code''}.
  In \bibinfo{booktitle}{{\em Proceedings of the 42Nd Annual {ACM}
  {SIGPLAN-SIGACT} Symposium on Principles of Programming Languages}} {\em
  (\bibinfo{series}{POPL '15})}. \bibinfo{publisher}{ACM},
  \bibinfo{address}{New York, NY, USA}, \bibinfo{pages}{111--124}.
\newblock
\showISBNx{9781450333009}
\showDOI{%
\url{https://doi.org/10.1145/2676726.2677009}}


\bibitem[\protect\citeauthoryear{Raychev, Vechev, and Yahav}{Raychev
  et~al\mbox{.}}{2014}]%
        {Raychev:2014}
\bibfield{author}{\bibinfo{person}{Veselin Raychev}, \bibinfo{person}{Martin
  Vechev}, {and} \bibinfo{person}{Eran Yahav}.}
  \bibinfo{year}{2014}\natexlab{}.
\newblock \showarticletitle{Code Completion with Statistical Language Models}.
  In \bibinfo{booktitle}{{\em Proceedings of the 35th ACM SIGPLAN Conference on
  Programming Language Design and Implementation}} {\em (\bibinfo{series}{PLDI
  '14})}. \bibinfo{publisher}{ACM}, \bibinfo{address}{New York, NY, USA},
  \bibinfo{pages}{419--428}.
\newblock
\showISBNx{978-1-4503-2784-8}
\showDOI{%
\url{https://doi.org/10.1145/2594291.2594321}}


\bibitem[\protect\citeauthoryear{Schilling}{Schilling}{2011}]%
        {Schilling2011-yf}
\bibfield{author}{\bibinfo{person}{Thomas Schilling}.}
  \bibinfo{year}{2011}\natexlab{}.
\newblock \showarticletitle{{Constraint-Free} Type Error Slicing}.
\newblock In \bibinfo{booktitle}{{\em Trends in Functional Programming}}.
  \bibinfo{publisher}{Springer Berlin Heidelberg}, \bibinfo{pages}{1--16}.
\newblock
\showISBNx{9783642320361, 9783642320378}
\showDOI{%
\url{https://doi.org/10.1007/978-3-642-32037-8\_1}}


\bibitem[\protect\citeauthoryear{Seidel and Jhala}{Seidel and Jhala}{2017}]%
        {Seidel2017-ko}
\bibfield{author}{\bibinfo{person}{Eric~L Seidel} {and} \bibinfo{person}{Ranjit
  Jhala}.} \bibinfo{year}{2017}\natexlab{}.
\newblock \bibinfo{title}{{A Collection of Novice Interactions with the {OCaml}
  {Top-Level} System}}.
\newblock   (\bibinfo{date}{June} \bibinfo{year}{2017}).
\newblock
\showDOI{%
\url{https://doi.org/10.5281/zenodo.806813}}


\bibitem[\protect\citeauthoryear{Seidel, Jhala, and Weimer}{Seidel
  et~al\mbox{.}}{2016}]%
        {Seidel2016-ul}
\bibfield{author}{\bibinfo{person}{Eric~L Seidel}, \bibinfo{person}{Ranjit
  Jhala}, {and} \bibinfo{person}{Westley Weimer}.}
  \bibinfo{year}{2016}\natexlab{}.
\newblock \showarticletitle{Dynamic Witnesses for Static Type Errors (or,
  Ill-typed Programs Usually Go Wrong)}. In \bibinfo{booktitle}{{\em
  Proceedings of the 21st {ACM} {SIGPLAN} International Conference on
  Functional Programming}} {\em (\bibinfo{series}{ICFP 2016})}.
  \bibinfo{publisher}{ACM}, \bibinfo{address}{New York, NY, USA},
  \bibinfo{pages}{228--242}.
\newblock
\showISBNx{9781450342193}
\showDOI{%
\url{https://doi.org/10.1145/2951913.2951915}}


\bibitem[\protect\citeauthoryear{Serrano and Hage}{Serrano and Hage}{2016}]%
        {Serrano2016-oo}
\bibfield{author}{\bibinfo{person}{Alejandro Serrano} {and}
  \bibinfo{person}{Jurriaan Hage}.} \bibinfo{year}{2016}\natexlab{}.
\newblock \showarticletitle{Type Error Diagnosis for Embedded {DSLs} by
  {Two-Stage} Specialized Type Rules}.
\newblock In \bibinfo{booktitle}{{\em Programming Languages and Systems}}.
  \bibinfo{publisher}{Springer Berlin Heidelberg}, \bibinfo{pages}{672--698}.
\newblock
\showISBNx{9783662494974, 9783662494981}
\showDOI{%
\url{https://doi.org/10.1007/978-3-662-49498-1\_26}}


\bibitem[\protect\citeauthoryear{Tip and Dinesh}{Tip and Dinesh}{2001}]%
        {Tip2001-qp}
\bibfield{author}{\bibinfo{person}{F Tip} {and} \bibinfo{person}{T~B Dinesh}.}
  \bibinfo{year}{2001}\natexlab{}.
\newblock \showarticletitle{A Slicing-based Approach for Locating Type Errors}.
\newblock \bibinfo{journal}{{\em ACM Trans. Softw. Eng. Methodol.\/}}
  \bibinfo{volume}{10}, \bibinfo{number}{1} (\bibinfo{date}{Jan.}
  \bibinfo{year}{2001}), \bibinfo{pages}{5--55}.
\newblock
\showISSN{1049-331X}
\showDOI{%
\url{https://doi.org/10.1145/366378.366379}}


\bibitem[\protect\citeauthoryear{Wand}{Wand}{1986}]%
        {Wand1986-nw}
\bibfield{author}{\bibinfo{person}{Mitchell Wand}.}
  \bibinfo{year}{1986}\natexlab{}.
\newblock \showarticletitle{Finding the Source of Type Errors}. In
  \bibinfo{booktitle}{{\em Proceedings of the 13th {ACM} {SIGACT-SIGPLAN}
  Symposium on Principles of Programming Languages}} {\em
  (\bibinfo{series}{POPL '86})}. \bibinfo{publisher}{ACM},
  \bibinfo{address}{New York, NY, USA}, \bibinfo{pages}{38--43}.
\newblock
\showDOI{%
\url{https://doi.org/10.1145/512644.512648}}


\bibitem[\protect\citeauthoryear{Wong and Debroy}{Wong and Debroy}{2009}]%
        {Wong2009-pd}
\bibfield{author}{\bibinfo{person}{W~Eric Wong} {and} \bibinfo{person}{Vidroha
  Debroy}.} \bibinfo{year}{2009}\natexlab{}.
\newblock \bibinfo{booktitle}{{\em A survey of software fault localization}}.
\newblock \bibinfo{type}{{T}echnical {R}eport} UTDCS-45-09.
  \bibinfo{institution}{University of Texas at Dallas}.
\newblock


\bibitem[\protect\citeauthoryear{Yang}{Yang}{1999}]%
        {Yang1999-yr}
\bibfield{author}{\bibinfo{person}{Jun Yang}.} \bibinfo{year}{1999}\natexlab{}.
\newblock \showarticletitle{Explaining Type Errors by Finding the Source of a
  Type Conflict}. In \bibinfo{booktitle}{{\em Selected Papers from the 1st
  Scottish Functional Programming Workshop}} {\em (\bibinfo{series}{SFP '99})}.
  \bibinfo{publisher}{Intellect Books}, \bibinfo{address}{Exeter, UK},
  \bibinfo{pages}{59--67}.
\newblock
\showISBNx{9781841500249}


\bibitem[\protect\citeauthoryear{Yoo, Harman, and Clark}{Yoo
  et~al\mbox{.}}{2013}]%
        {Yoo2013-rw}
\bibfield{author}{\bibinfo{person}{Shin Yoo}, \bibinfo{person}{Mark Harman},
  {and} \bibinfo{person}{David Clark}.} \bibinfo{year}{2013}\natexlab{}.
\newblock \showarticletitle{Fault Localization Prioritization: Comparing
  Information-theoretic and Coverage-based Approaches}.
\newblock \bibinfo{journal}{{\em ACM Trans. Softw. Eng. Methodol.\/}}
  \bibinfo{volume}{22}, \bibinfo{number}{3} (\bibinfo{date}{July}
  \bibinfo{year}{2013}), \bibinfo{pages}{19:1--19:29}.
\newblock
\showISSN{1049-331X}
\showDOI{%
\url{https://doi.org/10.1145/2491509.2491513}}


\bibitem[\protect\citeauthoryear{Zhang and Myers}{Zhang and Myers}{2014}]%
        {Zhang2014-lv}
\bibfield{author}{\bibinfo{person}{Danfeng Zhang} {and}
  \bibinfo{person}{Andrew~C Myers}.} \bibinfo{year}{2014}\natexlab{}.
\newblock \showarticletitle{Toward General Diagnosis of Static Errors}. In
  \bibinfo{booktitle}{{\em Proceedings of the 41st {ACM} {SIGPLAN-SIGACT}
  Symposium on Principles of Programming Languages}} {\em
  (\bibinfo{series}{POPL '14})}. \bibinfo{publisher}{ACM},
  \bibinfo{address}{New York, NY, USA}, \bibinfo{pages}{569--581}.
\newblock
\showISBNx{9781450325448}
\showDOI{%
\url{https://doi.org/10.1145/2535838.2535870}}


\end{thebibliography}
